\documentclass[aps,prd,11pt,notitlepage,nofootinbib,superscriptaddress,showkeys,showpacs]{revtex4-1}
\pdfoutput=1
\usepackage{amstext,amsmath,amssymb,amsfonts,bbm,euscript,color,dsfont}
\usepackage[latin1]{inputenc}
\usepackage{epsfig}
\usepackage{hyperref}
\usepackage{amsthm}
\usepackage{subfigure}
\usepackage{color}
\usepackage{multirow}
\usepackage{bbold}
\usepackage{tikz}
\usetikzlibrary{arrows,backgrounds}
\usepackage{verbatim} 
\usepackage{graphicx}
\usepackage{latexsym}

\begin{document}

\title{\textbf{\Large Study of a gauge invariant local composite fermionic field}}


\author{M.~A.~L.~Capri}\email{caprimarcio@gmail.com}
\affiliation{Universidade do Estado do Rio de Janeiro (UERJ),
Instituto de F\'{i}sica, Departamento de F\'{i}sica Te\'{o}rica,
Rua S\~ao Francisco Xavier 524, Maracan\~{a}, Rio de Janeiro, Brasil, CEP 20550-013}


\author{S.~P.~Sorella} \email{silvio.sorella@gmail.com}
\affiliation{Universidade do Estado do Rio de Janeiro (UERJ),
Instituto de F\'{i}sica, Departamento de F\'{i}sica Te\'{o}rica,
Rua S\~ao Francisco Xavier 524, Maracan\~{a}, Rio de Janeiro, Brasil, CEP 20550-013}

\author{R.~C.~Terin} \email{rodrigoterin3003@gmail.com}
\affiliation{Universidade do Estado do Rio de Janeiro (UERJ),
Instituto de F\'{i}sica, Departamento de F\'{i}sica Te\'{o}rica,
Rua S\~ao Francisco Xavier 524, Maracan\~{a}, Rio de Janeiro, Brasil, CEP 20550-013}
\affiliation{Sorbonne Universit\'{e}, CNRS, Laboratoire de Physique Th\'{e}orique\\ de la Mati\`{e}re Condens\'{e}e, LPTMC, F-75005 Paris, France}


\begin{abstract}
\noindent In this work, we study a gauge invariant local non-polynomial composite  spinor field in the  fundamental representation in order to establish its renormalizability. Similar studies were already done in the case of pure Yang-Mills theories where a local composite gauge invariant vector field was obtained and an invariant renormalizable mass term  could be introduced. Our model consists of a massive Euclidean Yang-Mills action with gauge group $SU(N)$ coupled to fermionic matter in the presence of an invariant  spinor composite field and quantized in the linear covariant gauges.  The whole set of Ward identities is analysed  and the algebraic proof of the renormalizability of the model is obtained to all orders in a loop expansion.  
\end{abstract}

\maketitle

\section{Introduction}

\noindent The  theoretical studies of quantum chromodynamics (QCD) in the low energy regime, where nonperturbative phenomena like confinement take place,  still lack a satisfactory understanding. \\


\noindent Besides quark confinement, let us underline that the gluon confinement is still a challenging, yet unsolved, issue. Several nonperturbative techniques based on the studies of the Dyson-Schwinger equations, functional renormalization group, Kugo-Ojima criterion, Gribov-Zwanziger approach and its refined version, have provided a fruitful ground for a better understanding of the behaviour of the two-point Landau gauge gluon correlation function in the infrared region, see 
\cite{Alkofer:2000wg,Cornwall:1981zr,Cornwall:2013zra,Binosi:2009qm,Aguilar:2008xm,Aguilar:2015bud,Tissier:2010ts,Tissier:2011ey,Fischer:2008uz,Fischer:2009tn,Weber:2011nw,Frasca:2007uz,Siringo:2015wtx,Kugo:1979gm,Kugo:1995km,Chaichian:2018cyv,Gribov:1977wm,Zwanziger:1989mf,Zwanziger:1993dh,Vandersickel:2012tz,Dudal:2007cw,Dudal:2008sp,Dudal:2011gd,Dudal:2019ing,Gao:2017uox}. The output of these investigations is in quite good agreement with the lattice data on the gluon propagator, which exhibit a violation of the reflection positivity \cite{Cucchieri:2004mf,Cucchieri:2007rg,Duarte:2016iko,Oliveira:2012eh,Dudal:2013yva}. This peculiar behaviour of the gluon propagator is commonly interpreted as a signal of gluon confinement, due to the impossibility of attaching a physical meaning to the gluon as an excitation of the spectrum of the theory. \\

\noindent In the present work, we pursue the investigation started in  \cite{Fiorentini:2016rwx,Capri:2017bfd}, by extending it to the matter sector through the introduction  of a gauge invariant composite non-polynomial spinor field. The latter will be included in the starting action together  with the dimension two gauge invariant operator $A_{\min }^{2}$, obtained by minimizing $A^2$ along the gauge orbit \cite{Zwanziger:1990tn,Dell'Antonio:1989jn,Dell'Antonio:1991xt,vanBaal:1991zw}, namely 
\begin{equation}
A_{\min }^{2} = {\mathrm min}_{\{u\}}\mathrm{Tr} \int d^4x \,A_{\mu }^{u}A_{\mu }^{u} \;, \qquad 
A_{\mu }^{u}= u^{\dagger}A_{\mu}u+\frac{i}{g}u^{\dagger}\partial_{\mu}u \;. \label{mn}
\end{equation}
The minimization procedure for $A_{\min }^{2}$ enables one to introduce a nonlocal gauge field  $A_{\mu }^{h}$ which turns out to be left invariant by gauge transformations, order by order in powers of the coupling $g$ \cite{Lavelle:1995ty}. Recently, a localization procedure for $A_{\mu }^{h}$ has been achieved by means of a localizing Stueckelberg-like field $\xi$ \cite{Fiorentini:2016rwx,Capri:2017bfd}. The resulting non-polynomial action has been proven to be renormalizable to all orders thanks to the existence of a Becchi-Rouet-Stora-Tyutin (BRST) nilpotent exact symmetry. \\

\noindent In the sequel, in analogy with the localized expression for $A_{\mu }^{h}$, we shall introduce its fermionic counterpart, {\it i.e.} we shall construct a local composite non-polynomial spinor field $\psi^{h}$ which is left invariant by the BRST transformations. \\

\noindent The introduction of such a BRST invariant spinor field $\psi^{h}$ might have several practical applications. Let us quote, for example, the framework devised in \cite{Capri:2014bsa} where a kind of effective {\it horizon function } for matter fields was employed within the Gribov-Zwanziger setup in order to obtain an analytic expression for the quark propagator in agreement with the available lattice data in Landau gauge, see \cite{Capri:2014bsa} and refs. therein. Another potential fruitful application can be figure out in relationship to the non-Abelian Landau-Khalatnikov-Fradkin transformations (LKF) along the path outlined in \cite{DeMeerleer:2018txc}, where the localized BRST invariant composite field  $A_{\mu }^{h}$ was employed  to interpolate the $n$-point  correlation functions of the gauge field $A_\mu$ 
between different gauges. In this case, the composite invariant field $\psi^{h}$ could allow to generalize the construction of \cite{DeMeerleer:2018txc} to the LKF transformations for correlation functions including spinor fields within a BRST invariant renormalizable framework.\\

\noindent The paper is organized as follows. In Sec.~\ref{inv_action}, we outline the construction of a local and BRST invariant action including the gauge invariant composite fields $A_{\mu }^{h}$ and $\psi^{h}$. In Sec.~\ref{ward_ids}, we determine the Ward identities fulfilled by the resulting quantized action. In Sec.~\ref{renorm}, we work out the most general counterterm and establish the renormalization factors for all fields, composite operators, external sources and parameters. In Sec.~\ref{concl}, we present our conclusions. In the appendix ~\ref{CPT}, we describe with details the $CPT$ symmetries.

\section{ Construction of a local and BRST invariant action}
\label{inv_action}
\noindent Let us begin with the construction of a complete local BRST invariant
action containing the gauge invariant operators $A_{\mu}^{h}$, $\psi^{h}$ and $\bar{\psi}^{h}=(\psi^{h})^{\dagger}\gamma_{4}$. The order by order gauge invariant operator $A_{\mu}^{h}$ is already known in the literature \cite{Lavelle:1995ty,Fiorentini:2016rwx}, being given by a highly nonlocal expression in the gauge field, namely
\begin{eqnarray}
A^{h}_{\mu}=\left(\delta_{\mu\nu}-\frac{\partial_{\mu}\partial_{\nu}}{\partial^{2}}\right)
\left(A_{\nu}-ig\left[\frac{\partial A}{\partial^{2}}\,,\,A_{\nu}\right]
+\frac{ig}{2}\left[\frac{\partial A}{\partial^{2}}\,,\,\partial_{\nu}\frac{\partial A}{\partial^{2}}\right]\right)+\mathcal{O}(A^{3})\,. \label{nah}
\end{eqnarray} 
The operators $\psi^{h}$ and $\bar{\psi}^{h}$ are the analogous fermionic counterpart of  $A^{h}_{\mu}$ and will be discussed later in the local formulation, in section \eqref{psi}. In order to
establish a local framework for $A_{\mu}^{h}$, we introduce an auxiliary localizing Stueckelberg-like field $\xi^{^{a}}$, see
Refs. \cite{Delbourgo:1986wz,Delbourgo:1987np,Lavelle:1995ty,Dragon:1996tk,Ruegg:2003ps}, as follows
\begin{eqnarray}
h & =e^{ig\xi} & =e^{ig\xi^{a}T^{a}}\,,
\label{hxi-1}
\end{eqnarray}
where the matrices $\ensuremath{\{T^{a}\}}$ are the generators of
the gauge symmetry group $\ensuremath{SU(N)}$. Then, one can define
$A_{\mu}^{h}$ as
\begin{eqnarray}
A_{\mu}^{h} & \equiv A_{\mu}^{h,a}\,T^{a} & =h^{\dagger}A_{\mu}h+\frac{i}{g}h^{\dagger}\partial_{\mu}h\,,
\label{local_Ah-1}
\end{eqnarray}
provided the transversality constraint, 
\begin{eqnarray}
\partial_{\mu} A^{h}_{\mu}=0\,,  \label{tr} 
\end{eqnarray}
is fulfilled. \\

\noindent Expanding \eqref{local_Ah-1} in power series of $\xi$, we get
a non-polynomial series, 
\begin{eqnarray}
A^{h,a}_{\mu} & = & A_{\mu}^{a}-D^{ab}_{\mu}(A)\xi^{b}-\frac{g}{2}f^{abc}\xi^{b}D^{cd}_{\mu}(A)\xi^{d}+\mathcal{O}(\xi^{3})\,,
\end{eqnarray}
with 
\begin{eqnarray}
D^{ab}_{\mu}(A) = \delta^{ab}-gf^{abc}A^{c}_{\mu}
\end{eqnarray}
being the covariant derivative in the adjoint representation of $SU(N)$. The transversality constraint \eqref{tr} turns out to be thus a constraint on the Stueckelberg field. Indeed, solving iteratively the condition $\partial_{\mu} A^{h}_{\mu}=0$ for $\xi$, one gets back the non-local expression \eqref{nah}, see \cite{Fiorentini:2016rwx,Capri:2017bfd}. \\

\noindent The gauge invariance of expression \eqref{local_Ah-1} becomes now manifest. It is easy to check in fact that $A_{\mu}^{h}$ is left invariant by 
\begin{eqnarray}
A_{\mu}^u&\,\,\,\to\,\,\,&u^{\dagger}A_{\mu}u+\frac{i}{g}u^{\dagger}\partial_{\mu}u\,,\nonumber\\
h&\to& u^{\dagger}h\,,\nonumber\\
h^{\dagger}&\to&h^{\dagger} u\,,
\end{eqnarray}
\begin{equation}
   (A_{\mu}^{h})^{u} = A_{\mu}^{h} \;, \label{gah}
\end{equation}
with $u=e^{-ig\omega}\in SU(N)$.\\
\noindent From the local expression for $A^{h}_{\mu}$ a local non-polynomial gauge invariant action  is written down as
\begin{eqnarray}
S_{\mathrm{inv}} & = & \int d^{4}x\,\Bigg[\frac{1}{4}F_{\mu\nu}^{a}F_{\mu\nu}^{a}+\tau^{a}\partial_{\mu}A_{\mu}^{h,a}+\bar{\eta}^{a}\partial_{\mu}D_{\mu}^{ab}(A^{h})\eta^{b}+\frac{m^{2}}{2}A_{\mu}^{h,a}A_{\mu}^{h,a}\nonumber \\
 &  & +i\bar{\psi}^{i,\alpha}(\gamma_{\mu})_{\alpha\beta}D_{\mu}^{ij}(A)\psi^{j,\beta}-m_{\psi}\bar{\psi}_{\alpha}^{i}\psi^{i,\alpha}\Bigg]\,,\label{act11l}
\end{eqnarray}
where
\begin{eqnarray}
F^{a}_{\mu\nu}&=& \partial_{\mu}A^{a}_{\nu}-\partial_{\nu}A^{a}_{\mu}+gf^{abc}A^{b}_{\mu}A^{c}_{\nu}
\end{eqnarray}
is the field strength and
\begin{eqnarray}
D_{\mu}^{ij}(A) & = & \delta^{ij}\partial_{\mu}-ig(T^{a})^{ij}A_{\mu}^{a}\,
\end{eqnarray}
is the covariant derivatives in the fundamental representation of $SU(N)$. The action $\ensuremath{S_{\rm inv}}$ provides a local
setup for the nonlocal operator $A_{\mu}^{h}$. The auxiliary field $\tau^{a}$ plays the role of a Lagrange multiplier, implementing the transversality condition for the gauge
invariant composite field $A^{h}_{\mu}$. The fields $(\bar \eta, \eta)$ are a pair  of anti-commuting fields needed to take into account the Jacobian arising from the  integration in the functional integral over the multiplier $\tau^{a}$. To some extent, these fields are analogous to the Faddeev-Popov ghosts which will be introduced later on by means of the gauge-fixing procedure. Finally, the last two terms of $S_{\rm inv}$ represent the matter sector characterized by the Dirac fields $(\psi,\bar{\psi})$, while $m_{\psi}$ is the corresponding  mass parameter \footnote{According to the notations adopted here, the Greek indices $\{\mu,\nu,\rho,\sigma\}$ are the vector indices of the Euclidean space, while the Greek indices $\{\alpha,\beta,\gamma,\delta\}$ are the Dirac spinor indices. The Latin indices $\{a,b,c,d,e\}$, running from $1$ to $N^{2}-1$, are the indices of the adjoint representation of $SU(N)$; while the Latin indices $\{i,j,k,l\}$, running from $1$ to $N$, represent the indices of the fundamental representation of the group. The Dirac gamma matrices $\gamma_{\mu}$ in Euclidean space are given by
$$
\gamma_{4}=\left(
\begin{tabular}{cc}
    $0$ & $\mathbb{1}$ \\
     $\mathbb{1}$&$0$ 
\end{tabular}\right)\,,
\qquad
\gamma_{k}=-i\left(
\begin{tabular}{cc}
    $0$ & $\sigma_{k}$ \\
     $-\sigma_{k}$&$0$ 
\end{tabular}\right)\,,\qquad
\gamma_{5}=\gamma_{4}\gamma_{1}\gamma_{2}\gamma_{3}=\left(
\begin{tabular}{cc}
    $\mathbb{1}$ & $0$ \\
     $0$&$-\mathbb{1}$ 
\end{tabular}\right)\,,
$$
where $\mathbb{1}$ is the $2\times2$ identity matrix, $k=1,2,3$ and $\sigma_{k}$ are the well known Pauli matrices. }.\\

\noindent As pointed out in details in \cite{Fiorentini:2016rwx,Capri:2017npq}, the action \eqref{act11l} displays deep differences with respect to the conventional non-renormalizable non-Abelian Stueckelberg action \cite{Delbourgo:1986wz,Delbourgo:1987np,Dragon:1996tk,Ruegg:2003ps}. The difference lies precisely in the transversality constraint \eqref{tr}, implemented in expression \eqref{act11l} through the fields $(\tau, \bar \eta, \eta)$. Condition \eqref{tr} follows directly from the minimization procedure for the operator $A^{2}_{\rm min}$. As such, it has a geometrical meaning while being responsible for a good ultraviolet behavior of the model which, unlike the case of the standard Stueckelberg action, enjoys in fact perturbative renormalizability  \cite{Fiorentini:2016rwx,Capri:2017npq}. \\

\noindent It is worth observing that the gauge invariant composite operator $\ensuremath{A_{\mu}^{h,a}A_{\mu}^{h,a}}$ has been introduced in the action \eqref{act11l} through  the free mass parameter $m^2$. Interestingly, this term plays an essential role in the construction of effective  models which reproduce the infrared behaviour of the Landau gluon propagator observed in lattice simulation to high accuracy, see \cite{Tissier:2010ts,Tissier:2011ey,Siringo:2015wtx}.\\

\noindent As mentioned before, the action $S_{\rm inv}$ is left invariant by the following infinitesimal gauge transformations:
\begin{eqnarray}
A^{a}_{\mu} &\,\,\,\to\,\,\,& A^{a}_{\mu} -D_{\mu}^{ab}(A)\omega^{b}\,,\nonumber \\
\tau^{a}&\to&\tau^{a}\,,\nonumber \\
\eta^{a}&\to&\eta^{a}\,,\nonumber \\
\bar{\eta}^{a}&\to&\bar{\eta}^{a}\,,\nonumber \\
\psi_{\alpha}^{i} & \to&\psi_{\alpha}^{i} +ig(T^{a})^{ij}\omega^{a}\psi_{\alpha}^{j}\,,\nonumber \\
\bar{\psi}_{\alpha}^{i} & \to &\bar{\psi}_{\alpha}^{i} -ig\bar{\psi}_{\alpha}^{j}(T^{a})^{ji}\omega^{a}\,,\nonumber \\
\xi^{a}&\to& \xi^{a}-\omega^{a}+\frac{g}{2}f^{abc}\omega^{b}\xi^{c}-\frac{g^{2}}{12}f^{amr}f^{mpq}\omega^{p}\xi^{q}\xi^{r}+\mathcal{O}(\xi^{3})\,,\label{gauge_transfs}
\end{eqnarray}
where $\omega^{a}$ is the infinitesimal parameter of the gauge transformation. The non-polynomial transformations of the Stueckelberg field $\xi^a$ can be evaluated order by order by means of  the transformations of $h^{ij}=(e^{ig\xi})^{ij}$ and $(h^{\dagger})^{ij}$:
\begin{eqnarray}
h^{ij}&\to& h^{ij}+ig\omega^{a}(T^{a})^{ik}h^{kj}\,,\nonumber\\
(h^{\dagger})^{ij}&\to&(h^{\dagger})^{ij}-ig\omega^{a}(h^{\dagger})^{ik}(T^{a})^{kj}\,.
\end{eqnarray}

\subsection{Gauge-fixing and BRST symmetry}
Being the action $S_{\rm inv}$, eq.\eqref{act11l}, a gauge invariant expression,  a gauge-fixing condition is required for its quantization. In the present work we will choose the so-called linear covariant gauge, which is in fact a general class of  covariant  gauges depending on a gauge parameter $\alpha$. Therefore, we shall  consider the following gauge-fixing term:
\begin{eqnarray}
S_{\rm FP} = S_{\rm inv}+S_{\rm gf}\,.
\label{FP_action}
\end{eqnarray}
with
\begin{eqnarray}
S_{\rm gf}=\int d^{4}x\, \left( ib^{a}\,\partial_{\mu}A^{a}_{\mu}+\frac{\alpha}{2}b^{a}b^{a}+\bar{c}^{a}\,\partial_{\mu}D^{ab}_{\mu}(A)c^{b}\right)\,,  \label{gff} 
\end{eqnarray}
where $(\bar c{^a}, c^a)$ are the Faddeev-Popov ghosts and  $b^{a}$ is the auxiliary Nakanishi-Lautrup field implementing the gauge condition, namely
\begin{eqnarray}
\frac{\delta S_{\rm gf}}{\delta b^{a}}=i\partial_{\mu}A^{a}_{\mu}+\alpha\,b^{a}=0\,.
\label{gauge-fixing}
\end{eqnarray}
In particular, setting  $\alpha=0$, the Landau gauge is recovered.\\

\noindent As is well known, with the introduction of the gauge-fixing term $S_{\rm gf}$, the gauge invariance of expression \eqref{act11l} is replaced by the nilpotent BRST transformations which, in the present case, take the form    
\begin{eqnarray}
sA^{a}_{\mu} &\,\,\,=\,\,\,& -D_{\mu}^{ab}(A)c^{b}\,,\nonumber \\
sc^{a}&=&\frac{g}{2}f^{abc}c^{b}c^{c}\,,\nonumber\\
s\bar{c}^{a}&=&ib^{a}\,,\nonumber\\
sb^{a}&=&0\,,\nonumber\\
s\tau^{a}&=&0\,,\nonumber \\
s\eta^{a}&=&0\,,\nonumber \\
s\bar{\eta}^{a}&=&0\,,\nonumber \\
s\psi_{\alpha}^{i} &=& ig(T^{a})^{ij}c^{a}\psi_{\alpha}^{j}\,,\nonumber \\
s\bar{\psi}_{\alpha}^{i} & = & -ig\bar{\psi}_{\alpha}^{j}(T^{a})^{ji}c^{a}\,,\nonumber \\
s\xi^{a}&=&g^{ab}(\xi)c^{b}\,,\label{BRST_transfs}
\end{eqnarray}
where,
\begin{equation}
    g^{ab}(\xi)=-\delta^{ab}+\frac{g}{2}f^{abc}\xi^{c}-\frac{g^{2}}{12}f^{aec}f^{ebd}\xi^{c}\xi^{d}+\mathcal{O}(\xi^{3})\,,
\end{equation}
and 
\begin{equation} 
s S_{\rm FP} =0 \;, \qquad s^2=0\;. 
\end{equation}
Evidently, we also have 
\begin{eqnarray}
sA^{h}_{\mu}=0\,.
\end{eqnarray}
Let us end this section by pointing out that the Stueckelberg field $\xi$ is a massless field, whose propagator behaves like $\langle \xi \xi\rangle_p \sim \alpha/p^4$, a feature which might eventually lead to undesired spurious infrared divergences in some Green functions. Though, as shown in \cite{Capri:2017npq}, the gauge fixing \eqref{gff} 
can be suitable modified to account for a renormalizable BRST invariant infrared regurlarization for the field $\xi$. More precisley, one can introduce a regularizing infrared mass $\mu^2$ through the exact BRST term 
\begin{eqnarray}
S_{\rm gf}^{\mu} & = & \int d^4x \;s \left ({\bar c}^a (\partial_\mu A^a_\mu - \mu^2 \xi^a)-i {\frac{\alpha}{2}} {\bar c}^a b^a \right) \nonumber \\
&=& \int d^{4}x\, \left( ib^{a}\,\partial_{\mu}A^{a}_{\mu}+\frac{\alpha}{2}b^{a}b^{a}+\bar{c}^{a}\,\partial_{\mu}D^{ab}_{\mu}(A)c^{b} -i\mu^2 b^a \xi^a + \mu^2 {\bar c}^a g^{ab}(\xi) c^b \right)\, \label{gff1} \;.  
\end{eqnarray}
It is easy to see  that the gauge parameter $\mu^2$ plays the role of an infrared regulator for the $\xi$ field, whose propagator gets now the infrared safe form $\langle \xi \xi\rangle_p \sim \alpha/(p^2+\mu^2)^2$. Moreover, as much as the parameter $\alpha$, also  $\mu^2$ appears in a BRST exact term, eq.\eqref{gff1}. As a consequence, it is a pure gauge parameter which will not affect the correlation functions of the gauge invariant operators, which are the object of the present work. Therefore, we shall proceeding by using the gauge fixing of eq.\eqref{gff}. 
\subsection{Introducing the local gauge invariant composite spinor  field $\psi^{h}$}\label{psi}
\noindent We are now ready to face the main task of the present work, namely: the introduction of the local non-polynomial gauge invariant spinor composite operator $\psi^{h}$ which, in terms of the auxiliary Stueckelberg field $\xi$, can be defined as 
\begin{eqnarray}
\psi^{h} \equiv  h^{\dagger}\psi\,,
\label{local_psih}
\end{eqnarray}
where $\psi$ is the Dirac field and $h$ is given by eq.\eqref{hxi-1}. As the Dirac field transforms as $\psi \rightarrow  u^{\dagger}\psi$ and $h^{\dagger}$ as $h^{\dagger}\rightarrow h^{\dagger} u$  for a finite gauge transformation $u$, it is immediate to realize that $\psi^{h}$ is gauge invariant. Of course, the same procedure can be done for the Dirac adjoint field $\bar{\psi}$, giving rise to the Dirac adjoint invariant composite field $\bar{\psi}^{h}\equiv\bar{\psi}h$.

Expanding in terms of the Stueckelberg field $\xi^{a}$, we have
\begin{eqnarray}
(\psi^{h})_{\alpha}^{i} & = & \psi_{\alpha}^{i}-ig\xi^{a}(T^{a})^{ij}\psi_{\alpha}^{j}-\frac{g^{2}}{2}\xi^{a}\xi^{b}(T^{a})^{ij}(T^{b})^{jk}\psi_{\alpha}^{k}+\mathcal{O}(\xi^{3})\,.
\label{psih_expansion}
\end{eqnarray}
In order to study the composite fields $(A^{h}_{\mu},\psi^{h},\bar{\psi}^{h})$ at the quantum level, we introduce them in the starting action  by means of BRST invariant external sources which will allow to define the renormalized functional generators for  the Green functions involving the insertions of the aforementioned composite operators \cite{Piguet:1995er}. Moreover,  besides the invariant composite fields $(A^{h}_{\mu},\psi^{h},\bar{\psi}^{h})$, there are some other composite fields related to the nonlinear BRST transformation of the fields $(A^a_\mu, c^a, \psi^i, \bar{\psi}^i, \xi^a)$, eqs.\eqref{BRST_transfs}, which require the introduction of the corresponding BRST invariant  source. \\

\noindent Therefore, taking into account the whole set of external sources, we write down the following complete action $\Sigma$:
\begin{eqnarray}
\Sigma&=&\int d^{4}x\,\bigg[\frac{1}{4}F_{\mu\nu}^{a}F_{\mu\nu}^{a}+\tau^{a}\partial_{\mu}A_{\mu}^{h,a}+\bar{\eta}^{a}\partial_{\mu}D_{\mu}^{ab}(A^{h})\eta^{b}+i\bar{\psi}^{i,\alpha}(\gamma_{\mu})_{\alpha\beta}D_{\mu}^{ij}(A)\psi^{j,\beta}\nonumber\\
&&
+ib^{a}\,\partial_{\mu}A^{a}_{\mu}+\frac{\alpha}{2}b^{a}b^{a}+\bar{c}^{a}\,\partial_{\mu}D^{ab}_{\mu}(A)c^{b}\bigg]\nonumber\\
&&
+\int d^{4}x\,\biggl[\Omega^{a}_{\mu}(sA^{a}_{\mu})
+L^{a}(sc^{a})+K^{a}(s\xi^{a})
+(s\bar{\psi})U+\bar{U}(s\psi)
+\mathcal{J}^{a}_{\mu}A^{h,a}_{\mu}
+\bar{\psi}^{h,i}_{\alpha}V^{i\alpha}
+\bar{V}^{i}_{\alpha}\psi^{h,i\alpha}\nonumber\\
&&
+\frac{J}{2}\,A^{h,a}_{\mu}A^{h,a}_{\mu}
+J_{\psi}\,\bar{\psi}\psi
+\Xi^{a}_{\mu}\,D^{ab}_{\mu}(A^{h})\eta^{b}
+\frac{\zeta}{2}\,J^{2}+\zeta'\,J_{\psi}^{2}J+\zeta''\,J_{\psi}^{4}\bigg]\,.
\label{sigma_action}
\end{eqnarray}
One sees that the sources $(\mathcal{J}^{a}_{\mu}, V,\bar{V})$ couple to the composite operators $(A^{h}_{\mu},\bar{\psi}^{h},\psi^{h})$, while $(\Omega^{a}_{\mu},L^{a},K^{a},U,\bar{U})$ to the nonlinear BRST transformations of the fields $(A^{a}_{\mu},c^{a},\xi^{a},\bar{\psi},\psi)$.  For later convenience, the invariant mass terms $(A^h_\mu A^h_\mu)$ and $(\bar \psi \psi)$ are also introduced through local sources $(J,J_{\psi})$. Moreover, it turns out to be useful to introduce also the composite field operator $D^{ab}_{\mu}(A^{h})\eta^{b}$ by means of its corresponding  source $\Xi^{a}_{\mu}$. Finally, The last three terms in eq.\eqref{sigma_action}, containing the free parameters $(\zeta,\zeta', \zeta'')$, are allowed by power counting. These terms are needed to reabsorb the ultraviolet divergences appearing in the correlation fuctions $\langle (A^hA^h)_x (A^hA^h)_y\rangle $, $\langle (A^hA^h)_x (A^hA^h)_y  ({\bar{\psi}} \psi)_z \rangle $ and $\langle ({\bar{\psi}} \psi)_x ({\bar{\psi}} \psi)_y({\bar{\psi}} \psi)_z({\bar{\psi}} \psi)_t \rangle$.\\\\ The BRST invariance of $\Sigma$ is automatically fulfilled by setting  
\begin{equation}
s(\Omega,L,K,U,\bar{U},\mathcal{J},V,\bar{V},\Xi, J,J_{\psi})=0\,.    
\end{equation}
Finally, the Faddeev-Popov action, eq.\eqref{FP_action}, is recovered from $\Sigma$ when the external sources $(\Omega,L,K,U,\bar{U},\mathcal{J},V,\bar{V},\Xi)$ vanish and  $(J,J_\psi)$ are set equal to:
\begin{equation}
J(x)\big|_{\rm phys}=m^{2}\,,\qquad J_{\psi}(x)\big|_{\rm phys}=m_{\psi}\,.
\label{phys_values}
\end{equation}
We are now ready to establish the symmetry content of the action  \eqref{sigma_action}, a topic which will be faced in the next section.

\section{Ward Identities}
\label{ward_ids}

\noindent In order to go ahead  with  the algebraic renormalization procedure \cite{Piguet:1995er}, we need first to establish the Ward identities fulfilled by the action $\Sigma$, eq.\eqref{sigma_action}, which we enlist below, namely:

\subsection{The Slavnov-Taylor identity}
\noindent The BRST invariance can be expressed as a functional identity known as the  Slavnov-Taylor identity, 
\begin{equation}
    \mathcal{S}(\Sigma)=0\,.\label{sti}
\end{equation}
The nonlinear operator $\mathcal{S}$ is given, for a general functional $\mathcal{F}$, by
\begin{equation} 
\mathcal{S}(\mathcal{F}) =\int d^{4}x\left(\frac{\delta\mathcal{F}}{\delta A_{\mu}^{a}}\frac{\delta\mathcal{F}}{\delta\Omega_{\mu}^{a}}+\frac{\delta\mathcal{F}}{\delta c^{a}}\frac{\delta\mathcal{F}}{\delta L^{a}}+\frac{\delta\mathcal{F}}{\delta\xi^{a}}\frac{\delta\mathcal{F}}{\delta K^{a}}+\frac{\delta\mathcal{F}}{\delta U_{\alpha}^{i}}\frac{\delta\mathcal{F}}{\delta\bar{\psi}^{i,\alpha}}+\frac{\delta\mathcal{F}}{\delta\bar{U}_{\alpha}^{i}}\frac{\delta\mathcal{F}}{\delta\psi^{i,\alpha}}+ib^{a}\frac{\delta\mathcal{F}}{\delta{\bar{c}}^{a}}\right)\,. 
\end{equation}
For further use, we also introduce the so-called linearized Slavnov-Taylor operator $\mathcal{B}_{\Sigma}$ \cite{Piguet:1995er},
\begin{eqnarray}
{\cal B}_{\Sigma} & = & \int d^{4}x\,\bigg(\frac{\delta\Sigma}{\delta\Omega_{\mu}^{a}}\frac{\delta}{\delta A_{\mu}^{a}}+\frac{\delta\Sigma}{\delta A_{\mu}^{a}}\frac{\delta}{\delta\Omega_{\mu}^{a}}+\frac{\delta\Sigma}{\delta U_{\alpha}^{i}}\frac{\delta}{\delta\bar{\psi}^{i,\alpha}}+\frac{\delta\Sigma}{\delta\bar{\psi}_{\alpha}^{i}}\frac{\delta}{\delta U^{i,\alpha}} +\frac{\delta\Sigma}{\delta\bar{U}_{\alpha}^{i}}\frac{\delta}{\delta\psi^{i,\alpha}}\nonumber \\
&&
+\frac{\delta\Sigma}{\delta\psi_{\alpha}^{i}}\frac{\delta}{\delta\bar{U}^{i,\alpha}}+\frac{\delta\Sigma}{\delta L^{a}}\frac{\delta}{\delta c^{a}}+\frac{\delta\Sigma}{\delta c^{a}}\frac{\delta}{\delta L^{a}}+\frac{\delta\Sigma}{\delta K^{a}}\frac{\delta}{\delta\xi^{a}}
+\frac{\delta\Sigma}{\delta\xi^{a}}\frac{\delta}{\delta K^{a}}+ib^{a}\frac{\delta}{\delta\bar{c}^{a}}\bigg)\,,\label{lst}
\end{eqnarray}
enjoying the property of nilpotency,
\begin{eqnarray}
{\cal B}_{\Sigma}{\cal B}_{\Sigma} & = & 0\,.
\end{eqnarray}

\subsection{The equation of motion of the Lagrange multiplier $b^{a}$}

\noindent The equation of motion of  the Lagrange multiplier $b^{a}$, {\it i.e.}   
\begin{eqnarray}
\frac{\delta\Sigma}{\delta b^{a}} & = & i\partial_{\mu}A_{\mu}^{a}+\alpha b^{a}\,.\label{eq:beq}
\end{eqnarray}
corresponding to the linear covariant gauge-fixing adopted here, has the meaning of a Ward identity \cite{Piguet:1995er}. This follows from the fact that the right-hand side of \eqref{eq:beq} is linear in the quantum fields. As such, it represents a linearly breaking term which is not affected by quantum corrections \cite{Piguet:1995er}. 

\subsection{The antighost equation}

\noindent The antighost field $\bar{c}^{a}$ combines with the source $\Omega^{a}_{\mu}$ to give rise to  the following identity
\begin{eqnarray}
\frac{\delta\Sigma}{\delta\bar{c}^{a}}+\partial_{\mu}\frac{\delta\Sigma}{\delta\Omega_{\mu}^{a}} & = & 0\,,\label{eq:agh}
\end{eqnarray}
known as the antighost equation \cite{Piguet:1995er}.

\subsection{The equation of $\ensuremath{\tau^{a}}$}

\noindent Analogously to the antighost equation, the equation of motion of the $\tau^{a}$ field and the variation of the action with respect to the source $\mathcal{J}^{a}_{\mu}$, yields the following identity:
\begin{eqnarray}
\frac{\delta\Sigma}{\delta\tau^{a}}-\partial_{\mu}\frac{\delta\Sigma}{\delta\mathcal{J}_{\mu}^{a}} & = & 0\,,\label{teq}
\end{eqnarray}
which we shall call  the  $\tau$-equation.

\subsection{The $\eta$-ghost equation}

\noindent Differentiating the action with respect to the ghost  $\eta^{a}$ and integrating on space-time, gives the Ward identity:
\begin{equation}
    \mathcal{G}_{\eta}^{a}(\Sigma)=0\,,
    \label{etw}
\end{equation}
where $\mathcal{G}_{\eta}^{a}$ is the linear operator:
\begin{eqnarray}
\mathcal{G}_{\eta}^{a}=\int d^{4}x\,\left(\frac{\delta}{\delta\eta^{a}}+gf^{abc}\bar{\eta}^{b}\frac{\delta}{\delta\tau^{c}}+gf^{abc}\Xi_{\mu}^{b}\frac{\delta}{\delta\mathcal{J}_{\mu}^{c}}\right)\,.
\end{eqnarray}

\subsection{The $\bar{\eta}$-antighost equation}

\noindent  Similar to the antighost equation, eq.\eqref{eq:agh}, is the  $\bar{\eta}$-antighost Ward identity, namely  
\begin{eqnarray}
\frac{\delta\Sigma}{\delta\bar{\eta}^{a}}-\partial_{\mu}\frac{\delta\Sigma}{\delta\Xi_{\mu}^{a}} & = & 0\,.
\end{eqnarray}
Notice that the presence of the composite field operator $D^{ab}_{\mu}(A^{h})\eta^{b}$, coupled to the source $\Xi^{a}_{\mu}$, is needed in order to establish this identity.

\subsection{The $\eta$-ghost number}

\noindent A ghost number can be assigned to the anti-commuting fields $(\bar \eta, \eta)$ and to the source $\Xi_{\mu}$, resulting in the following $\eta$-ghost number Ward identity
\begin{eqnarray}
\mathcal{Q}_{\eta}(\Sigma)= \int d^{4}x\,\left(\eta^{a}\frac{\delta\Sigma}{\delta\eta^{a}}-\bar{\eta}^{a}\frac{\delta\Sigma}{\delta\bar{\eta}^{a}}-\Xi_{\mu}^{a}\frac{\delta\Sigma}{\delta\Xi_{\mu}^{a}}\right) & = & 0\;.\label{ghn}
\end{eqnarray}

\subsection{The $c$-ghost number}

\noindent Analogously, we have also the usual $c$-ghost number in the Faddeev-Popov sector, expressed by 
\begin{eqnarray}
\mathcal{Q}_{c}(\Sigma)& = & \int d^{4}x\,\bigg(c^{a}\frac{\delta\Sigma}{\delta c^{a}}-\bar{c}^{a}\frac{\delta\Sigma}{\delta\bar{c}^{a}}-\Omega_{\mu}^{a}\frac{\delta\Sigma}{\delta\Omega_{\mu}^{a}}-2L^{a}\frac{\delta\Sigma}{\delta L^{a}}-K^{a}\frac{\delta\Sigma}{\delta K^{a}}\nonumber \\
 && -U^{i,\alpha}\frac{\delta\Sigma}{\delta U^{i,\alpha}}-\bar{U}^{i,\alpha}\frac{\delta\Sigma}{\delta\bar{U}^{i,\alpha}}\bigg)=0\,.
  \label{gn}
\end{eqnarray}

\subsection{The fermionic Ward identity}
\noindent Finally, similarly to the  $\eta$ and $c$-ghost numbers, we have  a Ward identity in the matter fermionic sector, given by 
\begin{eqnarray}
\mathcal{{Q}}_{\psi}(\Sigma) & = & \int d^{4}x\,\Bigg(\psi_{\alpha}^{i}\frac{\delta\Sigma}{\delta\psi^{i,\alpha}}-\bar{\psi}_{\alpha}^{i}\frac{\delta\Sigma}{\delta\bar{\psi}^{i,\alpha}}+U_{\alpha}^{i}\frac{\delta\Sigma}{\delta U^{i,\alpha}}-\bar{U}{}_{\alpha}^{i}\frac{\delta\Sigma}{\delta\bar{U}^{i,\alpha}}+V^{i}_{\alpha}\frac{\delta\Sigma}{\delta V^{i,\alpha}}-\bar{V}{}^{i}_{\alpha}\frac{\delta\Sigma}{\delta\bar{V}^{i,\alpha}}\Bigg) =0 \,.\nonumber \\
\label{u1}
\end{eqnarray}
The quantum numbers of all fields and sources are displayed in  Tables \ref{tbb1} and \ref{tbb2}, where we use the capital letter (C) for the commuting quantities and (A) for the anti-commuting ones. \\

\begin{table}
\begin{tabular}{|c|c|c|c|c|c|c|c|c|c|c|}
\hline 
Fields & $\ensuremath{A_{\mu}^{a}}$ & $\ensuremath{b^{a}}$ & $\ensuremath{c^{a}}$ & $\ensuremath{\bar{c}^{a}}$ & $\ensuremath{\tau^{a}}$ & $\ensuremath{\eta^{a}}$ & $\ensuremath{\bar{\eta}^{a}}$ & $\ensuremath{\xi^{a}}$ & $\bar{\psi}^{i,\alpha}$ & $\psi^{i,\alpha}$\tabularnewline
\hline 
\hline 
mass dimension $(d)$& $1$ & $2$ & $0$ & $2$ & $2$ & $0$ & $2$ & $0$ & $\frac{3}{2}$ & $\frac{3}{2}$\tabularnewline
\hline 
$c$-ghost number  & $0$ & $0$ & $1$ & $-1$ & $0$ & $0$ & $0$ & $0$ & $0$ & $0$\tabularnewline
\hline 
$\eta$-ghost number  & $0$ & $0$ & $0$ & $0$ & $0$ & $1$ & $-1$ & $0$ & $0$ & $0$\tabularnewline
\hline 
$\psi$-charge  & $0$ & $0$ & $0$ & $0$ & $0$ & $0$ & $0$ & $0$ & $-1$ & $1$\tabularnewline
\hline 
Nature & C & C & A & A & C & A & A & C & A & A\tabularnewline
\hline 
\end{tabular}
\caption{The quantum numbers of the fields} 
\label{tbb1}
\end{table}

\begin{table}
\begin{tabular}{|c|c|c|c|c|c|c|c|c|c|}
\hline 
Sources & $\Omega_{\mu}^{a}$ & $L^{a}$ & $K^{a}$ & $\mathcal{J}_{\mu}^{a}$ & $\Xi_{\mu}^{a}$ & $\bar{U}_{\alpha}^{i}$ & $U_{\alpha}^{i}$ & $\bar{V}_{\alpha}^{i}$ & $V_{\alpha}^{i}$\tabularnewline
\hline 
\hline 
mass dimension $(d)$ & $3$ & $4$ & $4$ & $3$ & $2$ & $\frac{5}{2}$ & $\frac{5}{2}$ & $\frac{5}{2}$ & $\frac{5}{2}$\tabularnewline
\hline 
$c$-ghost number  & $-1$ & $-2$ & $-1$ & $0$ & $0$ & $-1$ & $-1$ & $0$ & $0$\tabularnewline
\hline 
$\eta$-ghost number  & $0$ & $0$ & $0$ & $0$ & $-1$ & $0$ & $0$ & $0$ & $0$\tabularnewline
\hline 
$\psi$-charge  & $0$ & $0$ & $0$ & $0$ & $0$ & $-1$ & $1$ & $-1$ & $1$\tabularnewline
\hline 
Nature & A & C & A & C & C & C & C & A & A\tabularnewline
\hline 
\end{tabular}
\caption{The quantum numbers of the external sources}
\label{tbb2} 
\end{table} 

\noindent As we shall see, this set of Ward identities and the CPT symmetries are enough to prove the renormalizability of the action eq.\eqref{sigma_action} by means of the algebraic renormalization procedure \cite{Piguet:1995er}.

\section{Renormalization}
\label{renorm}
\noindent In the last section, we presented the set of Ward Identities that action $\Sigma$, eq.\eqref{sigma_action}, obeys. We turn now our attention to determine the most general  invariant counterterm  preserving such identities, which will enable us to prove the renormalizability of the starting action $\Sigma$  in the presence of the gauge invariant composite fields $(A^{h}_{\mu},\psi^{h},\bar{\psi}^{h})$.

\subsection{Determination of the most general counterterm}

\noindent In order to characterize the most general invariant counterterm, we follow the algebraic renormalization setup \cite{Piguet:1995er} and  perturb the action $\Sigma$,
\begin{equation}
    \Sigma\to\Sigma+\epsilon\,\Sigma_{\rm count}\,,
\end{equation}
in such a way that $\epsilon$ is an expansion parameter and $\Sigma_{\rm count}$ is,  
in agreement with the power-counting, an integrated local polynomial in the fields and sources with dimension four and vanishing $c$-ghost, $\eta$-ghost and $\psi$ quantum numbers.  Demanding that the perturbed action, $(\Sigma + \epsilon\,\Sigma_{\rm count})$, fulfills to the first order in the expansion parameter $\epsilon$  the same Ward identities of the action $\Sigma$, namely  
\begin{eqnarray}
\mathcal{S}(\Sigma+\epsilon\,\Sigma_{{\rm count}}) & = & 0+\mathcal{O}(\epsilon^{2})\,,\nonumber \\
\frac{\delta}{\delta b^{a}}(\Sigma+\epsilon\,\Sigma_{{\rm count}}) & = & i\partial_{\mu}A_{\mu}^{a}+\alpha\,b^{a}+\mathcal{O}(\epsilon^{2})\,,\nonumber \\
\left(\frac{\delta}{\delta\bar{c}^{a}}+\partial_{\mu}\frac{\delta}{\delta\Omega_{\mu}^{a}}\right)(\Sigma+\epsilon\,\Sigma_{{\rm count}}) & = & 0+\mathcal{O}(\epsilon^{2})\,,\nonumber \\
\left(\frac{\delta}{\delta{\tau}^{a}}-\partial_{\mu}\frac{\delta}{\delta\mathcal{J}_{\mu}^{a}}\right)(\Sigma+\epsilon\,\Sigma_{{\rm count}}) & = & 0+\mathcal{O}(\epsilon^{2})\,,\nonumber \\
\left(\frac{\delta}{\delta\bar{\eta}^{a}}+\partial_{\mu}\frac{\delta}{\delta\Xi_{\mu}^{a}}\right)(\Sigma+\epsilon\,\Sigma_{{\rm count}}) & = & 0+\mathcal{O}(\epsilon^{2})\,,\nonumber \\
\mathcal{G}_{\eta}^{a}(\Sigma+\epsilon\,\Sigma_{{\rm count}}) & = & 0+\mathcal{O}(\epsilon^{2})\,,\nonumber \\
\mathcal{Q}_{\eta}(\Sigma+\epsilon\,\Sigma_{{\rm count}}) & = & 0+\mathcal{O}(\epsilon^{2})\,,\nonumber \\
\mathcal{Q}_{c}(\Sigma+\epsilon\,\Sigma_{{\rm count}}) & = & 0+\mathcal{O}(\epsilon^{2})\,,\nonumber \\
\mathcal{Q}_{\psi}(\Sigma+\epsilon\,\Sigma_{{\rm count}}) & = & 0+\mathcal{O}(\epsilon^{2})\,,
\label{stp}
\end{eqnarray}
we obtain the following constraints on the local invariant counterterm  $\Sigma_{\rm count}$:
\begin{eqnarray}
{\cal B}_{\Sigma}\Sigma_{\rm count} & = & 0\,,\label{cc1}\\\cr
\frac{\delta\Sigma_{\rm count}}{\delta b^{a}} & = & 0\,,\label{cc2}\\\cr
\left(\frac{\delta}{\delta\bar{c}^{a}}+\partial_{\mu}\frac{\delta}{\delta\Omega_{\mu}^{a}}\right)\Sigma_{\rm count} & = & 0\,,\label{cc3}\\\cr
\left(\frac{\delta}{\delta\tau^{a}}-\partial_{\mu}\frac{\delta}{\delta\mathcal{J}_{\mu}^{a}}\right)\Sigma_{\rm count} & = & 0\,,\label{cc4}\\\cr
\left(\frac{\delta}{\delta\bar{\eta}^{a}}-\partial_{\mu}\frac{\delta}{\delta\Xi_{\mu}^{a}}\right)\Sigma_{\rm count} & = & 0\,,\label{cc6}\\\cr
\mathcal{G}^{a}_{\eta}(\Sigma_{\rm count}) & = & 0\,,\label{cc5}\\\cr
\mathcal{Q}_{\eta}(\Sigma_{\rm count}) & = & 0\,,\label{cc7}\\\cr
\mathcal{Q}_{c}(\Sigma_{\rm count})&=&0\,,\label{cc8}\\\cr
\mathcal{Q}_{\psi}(\Sigma_{\rm count})&=&0\,,\label{cc9}
\end{eqnarray}
where $\mathcal{B}_{\Sigma}$ in eq.\eqref{cc1} is the linearized Slavnov-Taylor operator, as defined in eq.\eqref{lst}. From the nilpotency of $\mathcal{B}_{\Sigma}$, it follows that the first condition, eq.\eqref{cc1}, implies that the counterterm
$\ensuremath{\Sigma_{\rm count}}$ belongs to the cohomology \cite{Piguet:1995er} of the operator
$\ensuremath{{\cal B}_{\Sigma}}$ in the space of the integrated local
polynomials in the fields, sources and parameters, of dimension four
and vanishing $c$-ghost,  $\eta$-ghost and  $\psi$ quantum numbers. \\

\noindent Making use of the general results on the cohomology of Yang-Mills theories, see \cite{Piguet:1995er},  it turns out that  $\ensuremath{\Sigma_{\rm count}}$ can be written
as 
\begin{eqnarray}
\Sigma_{\rm count} & = & \Delta_{\rm cohom}+{\cal B}_{\Sigma}\Delta^{\left(-1\right)}\,, \nonumber
\\
\Delta_{\rm cohom} & \neq & {\cal B}_{\Sigma} {\hat {\mathcal M}}^{(-1)} \;.
\label{pct}
\end{eqnarray}
for some local ${\hat {\mathcal M}}^{(-1)}$. The terms  $\ensuremath{\Delta_{\rm cohom}}$ and $\ensuremath{\Delta^{\left(-1\right)}}$ are, respectively,  the nontrivial and trivial solutions of \eqref{cc1}.  Also, according to the quantum numbers of the fields, $\ensuremath{\Delta^{\left(-1\right)}}$
is an integrated polynomial of dimension four, $c$-ghost number $-1$ and both vanishing 
$\ensuremath{\eta}$-ghost number and $\psi$-charge.\\

\noindent Taking into account the BRST invariance of $(A^h, \psi^h, {\bar \psi}^h, \eta, \bar \eta, \tau)$, it follows that the term  $\Delta_{\rm cohom}$ can be parametrized as 
\begin{eqnarray}
\Delta_{\rm cohom} & = & \int d^{4}x\,\Bigg[\frac{a_{0}}{4}\left(F_{\mu\nu}^{a}\right)^{2}+a_{1}\, J_{\psi}\bar{\psi}_{\alpha}^{i}\psi^{i,\alpha}+a_{2}\left(\partial_{\mu}A_{\mu}^{h,a}\right)\left(\partial_{\nu}A_{\nu}^{h,a}\right)+a_{3}\left(\partial_{\mu}A_{\nu}^{h,a}\right)\left(\partial_{\mu}A_{\nu}^{h,a}\right)\nonumber \\
 &&
 +a_{4}\,f^{abc}A^{h,a}_{\mu}A^{h,b}_{\nu}\,\partial_{\mu}A^{h,c}_{\nu}
 +a_{5}^{abcd}A_{\mu}^{h,a}A_{\mu}^{h,b}A_{\nu}^{h,c}A_{\nu}^{h,d}+\left(\partial_{\mu}\tau^{a}+\mathcal{J}_{\mu}^{a}\right)F_{\mu}^{a}\left(A,\xi\right)
 \nonumber\\
 &&
 +a_{6}\left(\partial_{\mu}\bar{\eta}^{a}+\Xi_{\mu}^{a}\right)\left(\partial_{\mu}\eta^{a}\right)
 +f^{abc}\left(\partial_{\mu}\bar{\eta}^{a}+\Xi_{\mu}^{a}\right)\eta^{b}G_{\mu}^{c}\left(A,\xi\right)
 +JI\left(A,\xi\right)
 +J_{\psi}^{2}\,I'\left(A,\xi\right)
\nonumber\\
&&
 +\bar{V}^{i,\alpha}\mathcal{F}_{\alpha}^{i}(\psi,\xi)+\bar{\mathcal{F}}_{\alpha}^{i}(\bar{\psi},\xi)V^{i,\alpha}
+a_{7}\,\frac{\zeta}{2}\,J^{2} 
+a_{8}\,\zeta'\,J^{2}_{\psi}J
+a_{9}\,\zeta''\,J^{4}_{\psi}\Bigg]\,,\label{e1}
\end{eqnarray}
where the coefficients $(a_{0}, a_{1}, a_{2}, a_{3}, a_{4}, a_{5}^{abcd}, a_{6},a_{7},a_8,a_9)$ are free constant dimensionless parameters and where we have already imposed the constraints \eqref{cc2}--\eqref{cc6} and \eqref{cc7}--\eqref{cc9}\footnote{We will leave the constraint \eqref{cc5} at the end.}. The quantities  $\ensuremath{F_{\mu}^{a}\left(A,\xi\right)}$, $\ensuremath{G_{\mu}^{a}\left(A,\xi\right)}$,
$\ensuremath{I\left(A,\xi\right)}$, $\ensuremath{I'\left(A,\xi\right)}$,
$\mathcal{F}_{\alpha}^{i}(\psi,\xi)$ and
$\bar{\mathcal{F}}_{\alpha}^{i}(\bar{\psi},\xi)$
stand for local  functionals which can be expanded in powers of $\xi^a $. \\
\noindent Requiring now that $\Delta_{\rm cohom}$ fulfills  the constraint \eqref{cc1}, we obtain 
\begin{eqnarray}
&&{\cal B}_{\Sigma}{F}^{a}_{\mu}(A,\xi)=s{F}^{a}_{\mu}(A,\xi) = 0\,,\label{sF}\\
&&{\cal B}_{\Sigma}{G}^{a}_{\mu}(A,\xi)=s{G}^{a}_{\mu}(A,\xi) = 0\,,\label{sG}\\
&&{\cal B}_{\Sigma}{I}(A,\xi)=s{I}(A,\xi) = 0\,,\label{sI}\\
&&{\cal B}_{\Sigma}{I'}(A,\xi)=s{I'}(A,\xi) = 0\,,\label{sI_prime}\\
&&{\cal B}_{\Sigma}\mathcal{F}_{\alpha}^{i}(\psi,\xi)=s\mathcal{F}_{\alpha}^{i}(\psi,\xi) = 0\,,\label{17}\\
&&{\cal B}_{\Sigma}\bar{\mathcal{F}}_{\alpha}^{i}(\psi,\xi)=s\bar{\mathcal{F}}_{\alpha}^{i}(\psi,\xi) =  0\,.\label{17barra}
\end{eqnarray}
The quantities  $F^{a}_{\mu}(A,\xi)$, $G^{a}_{\mu}(A,\xi)$, $I(A,\xi)$ and $I'(A,\xi)$ have been already worked out in \cite{Fiorentini:2016rwx,Capri:2017bfd}, being given  by 
\begin{eqnarray}
F^{a}_{\mu}(A,\xi)&=&a_{10}\,A^{h,a}_{\mu}\,,\\
G^{a}_{\mu}(A,\xi)&=&a_{11}\,A^{h,a}_{\mu}\,,\\
I(A,\xi)&=&\frac{a_{12}}{2}\,A^{h,a}_{\mu}A^{h,a}_{\mu}\,,\\
I'(A,\xi)&=&\frac{a_{13}}{2}\,A^{h,a}_{\mu}A^{h,a}_{\mu}\,, 
\end{eqnarray}
with $(a_{10}, a_{11}, a_{12}, a_{13})$  free coefficients. 
Let us turn thus to  the solutions of eq.\eqref{17} and eq.\eqref{17barra}.  Let us first  analyze  eq.\eqref{17}. According to Table\,\ref{tbb1}, the quantity  $\mathcal{F}^{i}_{\alpha}(\psi,\xi)$ has the quantum numbers $d=3/2$, vanishing ghost and $\eta$ number, $\psi$-charge equals to 1 and is of  anticommuting nature. It also has two indices: one related to the internal symmetry group in the fundamental representation of $SU(N)$,  $(i)$, and the other being a spinor one $(\alpha)$. Therefore, taking into account all these properties, we are led to consider the following expression 
\begin{eqnarray}
\mathcal{F}_{\alpha}^{i}(\psi,\xi) & = & \sigma_{\alpha\beta}^{ij}(\xi)\psi^{j\beta}\,,
\label{operator}
\end{eqnarray}
with $\sigma^{ij}_{\alpha\beta}(\xi)$ being a dimensionless quantity in the Stueckelberg field $\xi^{ij}=\xi^{a}T^{a,ij}$, given by
\begin{eqnarray}
\sigma_{\alpha\beta}^{ij}(\xi) & = & \sigma_{1}^{ij}(\xi)\delta_{\alpha\beta}+\sigma_{2}^{ij}(\xi)(\gamma_{4})_{\alpha\beta}+\sigma_{3}^{ij}(\xi)(\gamma_{1})_{\alpha\beta}+\sigma_{4}^{ij}(\xi)(\gamma_{2})_{\alpha\beta}+\sigma_{5}^{ij}(\xi)(\gamma_{3})_{\alpha\beta}\nonumber \\
 && +\sigma_{6}^{ij}(\xi)(\gamma_{5})_{\alpha\beta}+\sigma_{7}^{ij}(\xi)(\gamma_{4}\gamma_{5})_{\alpha\beta}+\sigma_{8}^{ij}(\xi)(\gamma_{1}\gamma_{5})_{\alpha\beta}+\sigma_{9}^{ij}(\xi)(\gamma_{2}\gamma_{5})_{\alpha\beta}\nonumber \\
 && +\sigma_{10}^{ij}(\xi)(\gamma_{3}\gamma_{5})_{\alpha\beta}+\sigma_{11}^{ij}(\xi)\left[\gamma_{4},\gamma_{1}\right]_{\alpha\beta}+\sigma_{12}^{ij}(\xi)\left[\gamma_{4},\gamma_{2}\right]_{\alpha\beta}+\sigma_{13}^{ij}(\xi)\left[\gamma_{4},\gamma_{3}\right]_{\alpha\beta}\nonumber \\
 && +\sigma_{14}^{ij}(\xi)\left[\gamma_{1},\gamma_{2}\right]_{\alpha\beta}+\sigma_{15}^{ij}(\xi)\left[\gamma_{1},\gamma_{3}\right]_{\alpha\beta}+\sigma_{16}^{ij}(\xi)\left[\gamma_{2},\gamma_{3}\right]_{\alpha\beta}\,.
 \label{sigma_expanded}
\end{eqnarray}
where $\gamma$'s are the Dirac matrices in Euclidean space.

\noindent Thanks to eq.\eqref{local_psih}, we can proceed as in \cite{Fiorentini:2016rwx,Capri:2017bfd} and  perform a convenient change of variables  in equation \eqref{operator}, which consists in the replacement  of $\psi_{\alpha}^{i}$  by the gauge invariant composite  field $\psi_{\alpha}^{h,i}$. Accordingly, the quantity $\sigma^{ij}_{\alpha\beta}(\xi)$ will be redefined, namely: 
\begin{eqnarray}
\mathcal{F}_{\alpha}^{i}(\psi,\xi) & = & \hat{\sigma}^{ij}_{\alpha\beta}(\xi)\psi^{h,j\beta}\,,
\label{newope}
\end{eqnarray}
with
\begin{eqnarray}
\hat{\sigma}_{\alpha\beta}^{ij}(\xi)&=& \sigma^{ik}_{\alpha\beta}(\xi)h^{kj}\nonumber\\
& = & \hat\sigma_{1}^{ij}(\xi)\delta_{\alpha\beta}
+\hat\sigma_{2}^{ij}(\xi)(\gamma_{4})_{\alpha\beta}
+\hat\sigma_{3}^{ij}(\xi)(\gamma_{1})_{\alpha\beta}
+\hat\sigma_{4}^{ij}(\xi)(\gamma_{2})_{\alpha\beta}
+\hat\sigma_{5}^{ij}(\xi)(\gamma_{3})_{\alpha\beta}\nonumber \\
 && 
 +\hat\sigma_{6}^{ij}(\xi)(\gamma_{5})_{\alpha\beta}
 +\hat\sigma_{7}^{ij}(\xi)(\gamma_{4}\gamma_{5})_{\alpha\beta}
 +\hat\sigma_{8}^{ij}(\xi)(\gamma_{1}\gamma_{5})_{\alpha\beta}
 +\hat\sigma_{9}^{ij}(\xi)(\gamma_{2}\gamma_{5})_{\alpha\beta}\nonumber \\
 && 
 +\hat\sigma_{10}^{ij}(\xi)(\gamma_{3}\gamma_{5})_{\alpha\beta}
 +\hat\sigma_{11}^{ij}(\xi)\left[\gamma_{4},\gamma_{1}\right]_{\alpha\beta}
 +\hat\sigma_{12}^{ij}(\xi)\left[\gamma_{4},\gamma_{2}\right]_{\alpha\beta}
 +\hat\sigma_{13}^{ij}(\xi)\left[\gamma_{4},\gamma_{3}\right]_{\alpha\beta}\nonumber \\
 && 
 +\hat\sigma_{14}^{ij}(\xi)\left[\gamma_{1},\gamma_{2}\right]_{\alpha\beta}
 +\hat\sigma_{15}^{ij}(\xi)\left[\gamma_{1},\gamma_{3}\right]_{\alpha\beta}
 +\hat\sigma_{16}^{ij}(\xi)\left[\gamma_{2},\gamma_{3}\right]_{\alpha\beta}\,,
 \label{bilinear}
\end{eqnarray}
where ${\hat \sigma}_{p}^{ij}(\xi)$, with $p=1,..,16$, are power series in $\xi^a$. \\
\noindent Let us now look again at the constraint \eqref{17}, {\it i.e.}: 
\begin{eqnarray}
{\cal B}_{\Sigma}\mathcal{F}_{\alpha}^{i}(\psi,\xi) & = & \int d^{4}x\,\Bigg(\frac{\delta\Sigma}{\delta\bar{U}_{\alpha}^{i}}\frac{\delta\mathcal{F}_{\alpha}^{i}(\psi,\xi)}{\delta\psi^{i,\alpha}}+\frac{\delta\Sigma}{\delta K^{k}}\frac{\delta\mathcal{F}_{\alpha}^{i}(\psi,\xi)}{\delta\xi^{k}}\Bigg)\nonumber \\
 & = & \int d^{4}x\,\frac{\delta\Sigma}{\delta K^{a}}\frac{\partial\hat{\sigma}^{ij}_{\alpha\beta}(\xi)}{\partial\xi^{a}}\psi^{h,j\beta}\nonumber \\
 & = & \int d^{4}x\, g^{ab}\left(\xi\right)c^{b}\frac{\partial\hat{\sigma}^{ij}_{\alpha\beta}(\xi)}{\partial\xi^{a}}\psi^{h,j\beta}
 \,\,=\,\, 0\,,
\end{eqnarray}
which immediately gives
\begin{eqnarray}
\frac{\partial\hat{\sigma}_{\alpha\beta}^{ij}(\xi)}{\partial\xi^{a}} =0\,.
\end{eqnarray}
Moreover, from the so-called discrete symmetries: {\it parity}, {\it time-reversal} and {\it charge conjugation}, we have the following result \footnote{For instance, see more details in appendix ~\ref{CPT}.},
\begin{eqnarray}
\hat{\sigma}_{1}^{ij}(\xi) = a_{14}\,\delta^{ij}\,,\qquad \hat{\sigma}_{n}^{ij}(\xi)=0 \,,
\end{eqnarray}
where $a_{14}$ is a constant and $n=2,\dots,16$. Thus, we finally conclude that the most general expression for $\mathcal{F}_{\alpha}^{i}(\psi,\xi)$ is given by
\begin{eqnarray}
\mathcal{F}_{\alpha}^{i}(\psi,\xi) & = & a_{14}\,\psi_{\alpha}^{h,i}\,.
\end{eqnarray}
The same procedure can be repeated for $\bar{\mathcal{F}}_{\alpha}^{i}(\bar{\psi},\xi)$,  yielding:
\begin{eqnarray}
\bar{\mathcal{F}}_{\alpha}^{i}(\bar{\psi},\xi) & = & a_{15}\,\bar{\psi}_{\alpha}^{h,i}\,,
\end{eqnarray}
with $a_{15}$ being a constant. Therefore, for the most general expression for $\Delta_{\rm cohom}$ we get
\begin{eqnarray}
\Delta_{\rm cohom} & = & \int d^{4}x\,\Bigg[\frac{a_{0}}{4}\left(F_{\mu\nu}^{a}\right)^{2}+a_{1}\, J_{\psi}\bar{\psi}_{\alpha}^{i}\psi^{i,\alpha}+a_{2}\left(\partial_{\mu}A_{\mu}^{h,a}\right)\left(\partial_{\nu}A_{\nu}^{h,a}\right)+a_{3}\left(\partial_{\mu}A_{\nu}^{h,a}\right)\left(\partial_{\mu}A_{\nu}^{h,a}\right)\nonumber \\
 &&
 +a_{4}\,f^{abc}A^{h,a}_{\mu}A^{h,b}_{\nu}\,\partial_{\mu}A^{c}_{\nu}
 +a_{5}^{abcd}A_{\mu}^{h,a}A_{\mu}^{h,b}A_{\nu}^{h,c}A_{\nu}^{h,d}+a_{10}\,\left(\partial_{\mu}\tau^{a}+\mathcal{J}_{\mu}^{a}\right)A^{h,a}_{\mu}
 \nonumber\\
 &&
 +a_{6}\left(\partial_{\mu}\bar{\eta}^{a}+\Xi_{\mu}^{a}\right)\left(\partial_{\mu}\eta^{a}\right)
 +a_{11}\,f^{abc}\left(\partial_{\mu}\bar{\eta}^{a}+\Xi_{\mu}^{a}\right)\eta^{b}A_{\mu}^{h,c}
 +\frac{a_{12}}{2}\,J\,A^{h,a}_{\mu}A^{h,a}_{\mu}
 +\frac{a_{13}}{2}\,J_{\psi}^{2}\,A^{h,a}_{\mu}A^{h,a}_{\mu}
 \nonumber\\
 &&
 +a_{14}\,\bar{V}^{i,\alpha}\psi^{h,i}_{\alpha}+a_{15}\,\bar{\psi}_{\alpha}^{h,i}(\bar{\psi},\xi)V^{i,\alpha}
+a_{7}\,\frac{\zeta}{2}\,J^{2} 
+a_{8}\,\zeta'\,J^{2}_{\psi}J
+a_{9}\,\zeta''\,J^{4}_{\psi}\Bigg]\,.
\end{eqnarray}

\noindent Let us turn now  to the trivial part of the cohomology of $\mathcal{B}_{\Sigma}$, {\it i.e.} to $\Delta^{(-1)}$.  Taking into account the quantum numbers of the fields and sources given by Tables \ref{tbb1} and \ref{tbb2},  $\Delta^{(-1)}$ can be written as: 
\begin{eqnarray}
\Delta^{(-1)} & = & \int d^{4}x\;\Bigg[f_{1}^{ab}\left(\xi,\alpha\right)\xi^{a}K^{b}+f_{2}^{ab}\left(\xi,\alpha\right)L^{a}c^{b}+f_{3}^{ab}\left(\xi,\alpha\right)\xi^{a}\left(\partial_{\mu}\Omega_{\mu}^{b}\right)\nonumber \\
 &  & +f_{4}^{ab}\left(\xi,\alpha\right)\left(\partial_{\mu}\xi^{a}\right)\Omega_{\mu}^{b}+f_{5}^{ab}\left(\xi,\alpha\right)A_{\mu}^{a}\Omega_{\mu}^{b}+f_{6}^{ab}\left(\xi,\alpha\right)A_{\mu}^{a}\left(\partial_{\mu}\bar{c}^{b}\right)\nonumber \\
 &  & +f_{7}^{ab}\left(\xi,\alpha\right)\left(\partial_{\mu}A_{\mu}^{a}\right)\bar{c}^{b}+f_{8}^{ab}\left(\xi,\alpha\right)\left(\partial_{\mu}\xi^{a}\right)\left(\partial_{\mu}\bar{c}^{b}\right)+f_{9}^{ab}\left(\xi,\alpha\right)\xi^{a}\left(\partial^{2}\bar{c}^{b}\right)\nonumber \\
 &  & +f_{10}^{ab}\left(\xi,\alpha\right)\bar{c}^{a}b^{b}+f_{11}^{ab}\left(\xi,\alpha\right)\bar{c}^{a}\tau^{b}+f_{12}^{abc}\left(\xi,\alpha\right)\bar{\eta}^{a}\eta^{b}\bar{c}^{c}+f_{13}^{abc}\left(\xi,\alpha\right)\bar{c}^{a}\bar{c}^{b}c^{c}\nonumber \\
 && +f_{14}^{ij}\left(\xi,\alpha\right)\bar{\psi}^{i,\alpha}U_{\alpha}^{j}+f^{ij}_{15}\left(\xi,\alpha\right)\bar{U}{}^{i,\alpha}\psi_{\alpha}^{j}\Bigg]\,,
\end{eqnarray}
with the arbitrary quantities $(f_{1}, ..., f_{15})$ being functionals of the field $\xi$ and of the gauge parameter $\alpha$\footnote{The terms $f^{ij}_{14}\,\bar{\psi}^{i,\alpha}U_{\alpha}^{j}$ and $f^{ij}_{15}\,\bar{U}^{i,\alpha}\psi_{\alpha}^{j}$ could be written in an even more general form being replaced by $f^{ij\alpha\beta}_{14}\,\bar{\psi}^{i}_{\alpha}U_{\beta}^{j}$ and $f^{ij\alpha \beta}_{15}\,\bar{U}{}^{i}_{\alpha}\psi_{\beta}^{j}$, with $f^{ij\alpha\beta}_{14}$ and $f^{ij\alpha\beta}_{15}$ expanded in the basis $\{\mathbb{1}, \gamma_{\mu},\gamma_{5},\gamma_{5}\gamma_{\mu},[\gamma_{\mu},\gamma_{\nu}]\}$ of the $16$ independent $4\times4$ matrices, as in the construction of $\sigma^{ij}_{\alpha\beta}$ in eq.\eqref{sigma_expanded}. But, applying the discrete symmetries (parity, time-reversal and charge conjugation) one can find that $f_{14}^{ij\alpha\beta}=f_{14}^{ij}\,\delta^{\alpha\beta}$ and $f_{15}^{ij\alpha\beta}=f_{15}^{ij}\,\delta^{\alpha\beta}$. }. \\

\noindent Imposing the constraints \eqref{cc2}\textendash\eqref{cc6}, $\Delta^{(-1)}$ becomes
\begin{eqnarray}
\Delta^{(-1)} & = & \int d^{4}x\;\Bigg[f_{1}^{ab}\left(\xi,\alpha\right)\xi^{a}K^{b}+f_{2}^{ab}\left(\xi,\alpha\right)L^{a}c^{b}+f_{3}^{ab}\left(\xi,\alpha\right)\xi^{a}\partial_{\mu}\left(\Omega_{\mu}^{b}+\partial_{\mu}\bar{c}^{b}\right)\nonumber \\
 &  & +f_{4}^{ab}\left(\xi,\alpha\right)\left(\partial_{\mu}\xi^{a}\right)\left(\Omega_{\mu}^{b}+\partial_{\mu}\bar{c}^{b}\right)+f_{5}^{ab}\left(\xi,\alpha\right)A_{\mu}^{a}\left(\Omega_{\mu}^{b}+\partial_{\mu}\bar{c}^{b}\right)\nonumber \\
 &  & +f_{14}^{ij}\left(\xi,\alpha\right)\bar{\psi}^{i,\alpha}U_{\alpha}^{j}+f_{15}^{ij}\left(\xi,\alpha\right)\bar{U}{}^{i,\alpha}\psi_{\alpha}^{j}\Bigg]\,.
 \label{25a}
\end{eqnarray}
We are left thus with the remaining constraint \eqref{cc5}. Nevertheless, observing that 
$\mathcal{B}_{\Sigma}\Delta^{(-1)}$ is independent from  $(\eta^{a},\tau,\mathcal{J}^{a}_{\mu})$, {\it i.e.} $ {\mathcal{G}^{a}_{\eta}}\left(\mathcal{B}_{\Sigma}\Delta^{(-1)}\right)=0 $, 
it turns out that
\begin{equation}
    \mathcal{G}^{a}_{\eta}(\Delta_{\rm cohom})=0\,,
\end{equation}
giving the condition 
\begin{equation} 
a_{11}=ga_{10}    \;, 
\end{equation} 
so that 
\begin{eqnarray}
\Delta_{\rm cohom} & = & \int d^{4}x\,\Bigg\{\frac{a_{0}}{4}\left(F_{\mu\nu}^{a}\right)^{2}+a_{1}\, J_{\psi}\bar{\psi}_{\alpha}^{i}\psi^{i,\alpha}+a_{2}\left(\partial_{\mu}A_{\mu}^{h,a}\right)\left(\partial_{\nu}A_{\nu}^{h,a}\right)+a_{3}\left(\partial_{\mu}A_{\nu}^{h,a}\right)\left(\partial_{\mu}A_{\nu}^{h,a}\right)\nonumber \\
 &&
 +a_{4}\,f^{abc}A^{h,a}_{\mu}A^{h,b}_{\nu}\,\partial_{\mu}A^{c}_{\nu}
 +a_{5}^{abcd}A_{\mu}^{h,a}A_{\mu}^{h,b}A_{\nu}^{h,c}A_{\nu}^{h,d}+a_{10}\,\bigg[\left(\partial_{\mu}\tau^{a}+\mathcal{J}_{\mu}^{a}\right)A^{h,a}_{\mu}
 \nonumber\\
 &&
 +gf^{abc}\left(\partial_{\mu}\bar{\eta}^{a}+\Xi_{\mu}^{a}\right)\eta^{b}A_{\mu}^{h,c}\bigg]
 +a_{6}\left(\partial_{\mu}\bar{\eta}^{a}+\Xi_{\mu}^{a}\right)\left(\partial_{\mu}\eta^{a}\right)
  +\frac{a_{12}}{2}\,J\,A^{h,a}_{\mu}A^{h,a}_{\mu}
 +\frac{a_{13}}{2}\,J_{\psi}^{2}\,A^{h,a}_{\mu}A^{h,a}_{\mu}
 \nonumber\\
 &&
 +a_{14}\,\bar{V}^{i,\alpha}\psi^{h,i}_{\alpha}
 +a_{15}\,\bar{\psi}_{\alpha}^{h,i}V^{i,\alpha}
+a_{7}\,\frac{\zeta}{2}\,J^{2} 
+a_{8}\,\zeta'\,J^{2}_{\psi}J
+a_{9}\,\zeta''\,J^{4}_{\psi}\Bigg\}\,.
\end{eqnarray}
Furthermore, following \cite{Fiorentini:2016rwx}, one realizes that setting the external sources $\mathbf{J}$, 
\begin{equation}
   \mathbf{J}\equiv K^{a},\mathcal{J}^{a}_{\mu},\Xi^{a}_{\mu}, J, J_{\psi},V^{i\alpha}, \bar{V}^{i\alpha}\,, 
\end{equation}
to zero in the action of eq.\eqref{sigma_action}, we obtain 
\begin{eqnarray}
\Sigma\big|_{\mathbf{J}=0}&=&\int d^{4}x\,\bigg[\frac{1}{4}F_{\mu\nu}^{a}F_{\mu\nu}^{a}+\tau^{a}\partial_{\mu}A_{\mu}^{h,a}+\bar{\eta}^{a}\partial_{\mu}D_{\mu}^{ab}(A^{h})\eta^{b}+i\bar{\psi}^{i,\alpha}(\gamma_{\mu})_{\alpha\beta}D_{\mu}^{ij}(A)\psi^{j,\beta}\nonumber\\
&&
+ib^{a}\,\partial_{\mu}A^{a}_{\mu}+\frac{\alpha}{2}b^{a}b^{a}+\bar{c}^{a}\,\partial_{\mu}D^{ab}_{\mu}(A)c^{b}\bigg]\nonumber\\
&&
+\int d^{4}x\,\biggl[\Omega^{a}_{\mu}(sA^{a}_{\mu})
+L^{a}(sc^{a})
+(s\bar{\psi})U+\bar{U}(s\psi)
\bigg]\,.
\label{sigma_action_reduced}
\end{eqnarray}
The fields $(\tau,\eta,\bar{\eta})$ can now be integrated out giving a unity \cite{Fiorentini:2016rwx,Capri:2017bfd}. As a consequence, the action above becomes that of the ordinary Yang-Mills theory in linear covariant
gauges with massless fermions. As such, when the sources  $\mathbf{J}$ are set to zero, the counterterm we are looking for must reduce to that  of the  standard Yang-Mills
theory in linear covariant gauges \cite{Piguet:1995er}. This implies that
\begin{equation}
    a_2=a_3=a_4=a_{5}^{abcd}=0\,,
\end{equation}
\begin{equation}
    a_6=-a_{10}\,,
\end{equation}
as well as 
\begin{eqnarray}
f_{2}^{ab}(\xi,\alpha) & = & \delta^{ab}\,a_{16}\,,\\
f_{5}^{ab}(\xi,\alpha) & = & \delta^{ab}\,a_{17}\,,\\
f_{14}^{ij}(\xi,\alpha)&=&\delta^{ij}\,a_{18}\,,\\
f_{15}^{ij}(\xi,\alpha)&=&\delta^{ij}\,a_{19}\,,
\end{eqnarray}
where $(a_{16}, a_{17}, a_{18}, a_{19})$ are free coefficient which may depend on the gauge parameter $\alpha$, due to the fact that they appear in the exact cohomology solution for 
the operator $\mathcal{B}_{\Sigma}$. In addition, we also have
\begin{eqnarray}
f_{3}^{ab}(\xi,\alpha)=f_{4}^{ab}(\xi,\alpha)& = & 0\,.
\end{eqnarray}
Therefore, for the expressions of  $\Delta_{\rm cohom}$ and $\Delta^{(-1)}$, we find, respectively
\begin{eqnarray}
\Delta_{\rm cohom} & =\! & \int d^{4}x\,\Bigg\{\frac{a_{0}}{4}\left(F_{\mu\nu}^{a}\right)^{2}+a_{1}\, J_{\psi}\bar{\psi}_{\alpha}^{i}\psi^{i,\alpha}
 +a_{10}\,\bigg[\left(\partial_{\mu}\tau^{a}+\mathcal{J}_{\mu}^{a}\right)A^{h,a}_{\mu}
 -(\partial_{\mu}\bar{\eta}^{a}+\Xi_{\mu}^{a})D^{ab}_{\mu}(A^{h})\eta^{b}\bigg]
\nonumber\\
&&
+\frac{a_{12}}{2}\,J\,A^{h,a}_{\mu}A^{h,a}_{\mu}
+\frac{a_{13}}{2}\,J_{\psi}^{2}\,A^{h,a}_{\mu}A^{h,a}_{\mu}
+a_{14}\,\bar{V}^{i,\alpha}\psi^{h,i}_{\alpha}
+a_{15}\,\bar{\psi}_{\alpha}^{h,i}V^{i,\alpha}
\nonumber\\
&&
+a_{7}\,\frac{\zeta}{2}\,J^{2} 
+a_{8}\,\zeta'\,J^{2}_{\psi}J
+a_{9}\,\zeta''\,J^{4}_{\psi}\Bigg\}\,,
\label{Delta}\\\cr
\Delta^{(-1)} & = \!&
\int d^{4}x\bigg[f_{1}^{ab}(\xi)\,\xi^{a}K^{b}
+a_{16}\,L^{a}c^{a}
+a_{17}\,(\Omega_{\mu}^{a}+\partial_{\mu}\bar{c}^{a})A_{\mu}^{a}
 +a_{18}\,\bar{\psi}^{i,\alpha}U_{\alpha}^{i}
 +a_{19}\,\bar{U}^{i,\alpha}\psi_{\alpha}^{i} \bigg]\,.
 \label{Delta_minus_one}
\end{eqnarray}
For later convenience, it is helpful to rewrite the whole counterterm, eqs.\eqref{Delta},\eqref{Delta_minus_one},  in parametric form. \\
\noindent From expressions eqs.\eqref{Delta}, \eqref{Delta_minus_one}, it is easy to show that:
\begin{eqnarray}
\Sigma_{\rm count}&=& \Delta_{\rm cohom}+\mathcal{B}_{\Sigma}\Delta^{(-1)}
\nonumber\\
&=&-a_{0}\,g^{2}\frac{\partial\Sigma}{\partial g^{2}}
+2a_{17}\,\alpha\frac{\partial\Sigma}{\partial\alpha}
+(a_{7}+2a_{12})\,\zeta\frac{\partial\Sigma}{\partial\zeta}
+[(a_{8}+2a_{1}+a_{12})\,\zeta'+a_{13}\,\zeta]\,\frac{\partial\Sigma}{\partial\zeta'}
\nonumber\\
&&
+[(a_{9}+4a_{1})\,\zeta''+a_{13}\,\zeta']\,\frac{\partial\Sigma}{\partial\zeta''}
+\int d^{4}x\,\bigg[
a_{1}\,J_{\psi}\frac{\delta\Sigma}{\delta J_{\psi}}
+(a_{12}\,J+a_{13}\,J^{2}_{\psi})\,\frac{\delta\Sigma}{\delta J}
\nonumber\\
&&
-a_{10}\,\tau^{a}\frac{\delta\Sigma}{\delta\tau^{a}}
+a_{10}\,\mathcal{J}^{a}_{\mu}\frac{\delta\Sigma}{\delta\mathcal{J}^{a}_{\mu}}
+a_{10}\,\bar{\eta}^{a}\frac{\delta\Sigma}{\delta\bar{\eta}^{a}}
-a_{10}\,\Xi^{a}_{\mu}\frac{\delta\Sigma}{\delta\Xi^{a}_{\mu}}
+a_{14}\,V^{i\alpha}\frac{\delta\Sigma}{\delta V^{i\alpha}}
\nonumber\\
&&
+a_{14}\,\bar{V}^{i\alpha}\frac{\delta\Sigma}{\delta \bar{V}^{i\alpha}}
+a_{16}\,L^{a}\frac{\delta\Sigma}{\delta L^{a}}
-a_{16}\,c^{a}\frac{\delta\Sigma}{\delta c^{a}}
+a_{17}\,A^{a}_{\mu}\frac{\delta\Sigma}{\delta A^{a}_{\mu}}
-a_{17}\,\Omega^{a}_{\mu}\frac{\delta\Sigma}{\delta\Omega^{a}_{\mu}}
\nonumber\\
&&
-a_{17}\,b^{a}\frac{\delta\Sigma}{\delta b^{a}}
-a_{17}\,\bar{c}^{a}\frac{\delta\Sigma}{\delta \bar{c}^{a}}
+a_{18}\,U^{i\alpha}\frac{\delta\Sigma}{\delta U^{i\alpha}}
+a_{18}\,\bar{U}^{i\alpha}\frac{\delta\Sigma}{\delta \bar{U}^{i\alpha}}
-a_{18}\,\psi^{i\alpha}\frac{\delta\Sigma}{\delta \psi^{i\alpha}}
\nonumber\\
&&
-a_{18}\,\bar{\psi}^{i\alpha}\frac{\delta\Sigma}{\delta \bar{\psi}^{i\alpha}}
+f_{1}^{ab}(\xi)\xi^{a}\frac{\delta\Sigma}{\delta\xi^{b}}
-\bigg(\frac{\partial f_{1}^{cb}(\xi)}{\partial \xi^{a}}\xi^{c}+f_{1}^{ab}(\xi)\bigg)K^{b}\frac{\delta\Sigma}{\delta K^{a}}\bigg]\,,
\label{parametric_form}
\end{eqnarray}
where we have considered, without loss of generality, $a_{14}=a_{15}$ and $a_{17}=a_{18}$, due to the fact the terms $\bar{V}^{i,\alpha}\psi^{h,i}_{\alpha}$ and $
\bar{\psi}_{\alpha}^{h,i}V^{i,\alpha}$  are  conjugates of each other, implying that $(\bar{\psi},\bar{U},\bar{V})$ renormalize as $({\psi},{U},{V})$. \\
\noindent In equation \eqref{parametric_form} we have also performed  the following rescaling\footnote{It is actually the same as replacing $S_{\rm YM}$ by $g^{-2}S_{\rm YM}$, so that the coupling constant $g$ does not appear anymore in the expression for $\Sigma$.}:
\begin{equation}
\left(A,\xi,\alpha,b,\tau,\mathcal{J},J\right)\to
\left(\frac{A}{g},\frac{\xi}{g},\frac{\alpha}{g^{2}},gb,g\tau,g\mathcal{J},g^{2}J\right)\,.
\end{equation}
The parametric form of the counterterm, eq.\eqref{parametric_form}, will significantly simplify the analysis of the renormalization factors, as we shall see in the next section.

\subsection{Analyzing the counterterm and renormalization factors}

\noindent We notice that $\Sigma_{\rm count}$, eqs.\eqref{Delta_minus_one},\eqref{parametric_form}, contains a large quantity of free coefficients, which need to be shown to correspond to renormalizations of the fields, coupling constant, masses, composite operators and external sources. In other words, we have to  prove that $\Sigma_{\rm count}$ can be reabsorbed into the original action by means of a redefinition of all its arguments, which will be generically denoted as $\{\Phi\}$, namely 
\begin{eqnarray}
\Sigma(\Phi)+\epsilon\Sigma_{\rm count}(\Phi) =  \Sigma(\Phi_{0})+\mathcal{O}(\epsilon^{2})\,,
\label{3}
\end{eqnarray}
where $\{\Phi_{0}\}$ stand for the so-called bare quantities. Thanks to the parametric form, eq.\eqref{parametric_form}, equation \eqref{3}  can be also rewritten as 
\begin{eqnarray}
\Sigma(\Phi_0)=(1+\epsilon\,\mathcal{R})\Sigma(\Phi)+\mathcal{O}(\epsilon^{2})\,,
\end{eqnarray}
where $\mathcal{R}$ denotes the operator
\begin{eqnarray}
\mathcal{R}
&=&-a_{0}\,g^{2}\frac{\partial}{\partial g^{2}}
+2a_{17}\,\alpha\frac{\partial}{\partial\alpha}
+(a_{7}+2a_{12})\,\zeta\frac{\partial}{\partial\zeta}
+[(a_{8}+2a_{1}+a_{12})\,\zeta'+a_{13}\,\zeta]\,\frac{\partial}{\partial\zeta'}
\nonumber\\
&&
+[(a_{9}+4a_{1})\,\zeta''+a_{13}\,\zeta']\,\frac{\partial}{\partial\zeta''}
+\int d^{4}x\,\bigg[
a_{1}\,J_{\psi}\frac{\delta}{\delta J_{\psi}}
+(a_{12}\,J+a_{13}\,J^{2}_{\psi})\,\frac{\delta}{\delta J}
\nonumber\\
&&
-a_{10}\,\tau^{a}\frac{\delta}{\delta\tau^{a}}
+a_{10}\,\mathcal{J}^{a}_{\mu}\frac{\delta}{\delta\mathcal{J}^{a}_{\mu}}
+a_{10}\,\bar{\eta}^{a}\frac{\delta}{\delta\bar{\eta}^{a}}
-a_{10}\,\Xi^{a}_{\mu}\frac{\delta}{\delta\Xi^{a}_{\mu}}
+a_{14}\,V^{i\alpha}\frac{\delta}{\delta V^{i\alpha}}
\nonumber\\
&&
+a_{14}\,\bar{V}^{i\alpha}\frac{\delta}{\delta \bar{V}^{i\alpha}}
+a_{16}\,L^{a}\frac{\delta}{\delta L^{a}}
-a_{16}\,c^{a}\frac{\delta}{\delta c^{a}}
+a_{17}\,A^{a}_{\mu}\frac{\delta}{\delta A^{a}_{\mu}}
-a_{17}\,\Omega^{a}_{\mu}\frac{\delta}{\delta\Omega^{a}_{\mu}}
\nonumber\\
&&
-a_{17}\,b^{a}\frac{\delta}{\delta b^{a}}
-a_{17}\,\bar{c}^{a}\frac{\delta}{\delta \bar{c}^{a}}
+a_{18}\,U^{i\alpha}\frac{\delta}{\delta U^{i\alpha}}
+a_{18}\,\bar{U}^{i\alpha}\frac{\delta}{\delta \bar{U}^{i\alpha}}
-a_{18}\,\psi^{i\alpha}\frac{\delta}{\delta \psi^{i\alpha}}
\nonumber\\
&&
-a_{18}\,\bar{\psi}^{i\alpha}\frac{\delta}{\delta \bar{\psi}^{i\alpha}}
+f_{1}^{ab}(\xi)\xi^{a}\frac{\delta}{\delta\xi^{b}}
-\bigg(\frac{\partial f_{1}^{cb}(\xi)}{\partial \xi^{a}}\xi^{c}+f_{1}^{ab}(\xi)\bigg)K^{b}\frac{\delta}{\delta K^{a}}\bigg]\,.
\label{R_operator}
\end{eqnarray}
This means that the quantities $\{\Phi\}$ and $\{\Phi_0\}$ are related by
\begin{eqnarray}
\Phi_{0}& = & (1+\epsilon\mathcal{R})\Phi+\mathcal{O}(\epsilon^{2})\,.
\label{renorms}
\end{eqnarray}
Defining the renormalization factors $\{Z\}$ as
\begin{eqnarray}
&
A_{0} =  Z_{A}^{1/2}\,A\,,\qquad b_{0}=Z_{b}^{1/2}\,b\,,\qquad
c_{0}=Z_{c}^{1/2}\,c\,,\qquad
\bar{c}_{0}=Z_{\bar{c}}^{1/2}\,\bar{c}\,,
&
\nonumber\\
&
\eta_{0}=Z_{\eta}^{1/2}\,\eta\,,\qquad
\bar{\eta}_{0}=Z_{\bar{\eta}}^{1/2}\,\bar{\eta}\,,\qquad
\tau_{0}=Z_{\tau}^{1/2}\,\tau\,,\qquad
\xi^{a}_{0}=Z^{ab}_{\xi}(\xi)\,\xi^{b}\,,
&
\nonumber\\
&
\psi_{0}=Z_{\psi}^{1/2}\,\psi\,,\qquad
\bar{\psi}_{0}=Z_{\bar{\psi}}^{1/2}\,\bar{\psi}\,,\qquad
U_{0}=Z_{U}\,U\,,\qquad
\bar{U}_{0}=Z_{\bar{U}}\,\bar{U}\,,
&
\nonumber\\
&
V_{0}=Z_{V}\,V\,,\qquad
\bar{V}_{0}=Z_{\bar{V}}\,\bar{V}\,,\qquad
L_{0}=Z_{L}\,L\,,\qquad
\Omega_{0}=Z_{\Omega}\,\Omega\,,\qquad
&
\nonumber\\
&
g_{0}=Z_{g}\,g\,,\qquad
\alpha_{0}=Z_{\alpha}\,\alpha\,,\qquad
\mathcal{J}_{0}=Z_{\mathcal{J}}\,\mathcal{J}\,,\qquad
K^{a}_{0}=Z^{ab}_{K}(\xi)\,K^{b}
&
\end{eqnarray}
and
\begin{equation}
    \left(
    \begin{tabular}{c}
    $J_{0}$\cr
    $J^{2}_{\psi,0}$
    \end{tabular}
    \right)
    =
    \left(
    \begin{tabular}{c|c}
    $Z_{J}$&$Z_{J,J_{\psi}}$\cr
    \hline
    0&$Z^{2}_{J_{\psi}}$
    \end{tabular}
    \right)
    \left(
    \begin{tabular}{c}
    $J$\cr
    $J^{2}_{\psi}$
    \end{tabular}
    \right)\,,\qquad
    \left(
    \begin{tabular}{c}
    $\zeta_{0}$\cr
    $\zeta'_{0}$\cr
    $\zeta''_{0}$
    \end{tabular}
    \right)=
        \left(
    \begin{tabular}{c|c|c}
    $Z_{\zeta}$&$0$&$0$\cr
    \hline
    $Z_{\zeta'\zeta}$&$Z_{\zeta'}$&$0$\cr
    \hline
    $0$&$Z_{\zeta''\zeta'}$&$Z_{\zeta''}$
    \end{tabular}
    \right)
    \left(
    \begin{tabular}{c}
    $\zeta$\cr
    $\zeta'$\cr
    $\zeta''$
    \end{tabular}
    \right)\,,
\end{equation}
we have, from eq.\eqref{renorms}, that
\begin{eqnarray}
&\displaystyle
Z_{g}=1-\epsilon\,\frac{a_{0}}{2}\,,\qquad
Z_{A}^{1/2}=1+\epsilon\,a_{17}\,,\qquad
Z_{\bar{\eta}}^{1/2}=1+\epsilon\,a_{10}\,,\qquad
Z_{\eta}^{1/2}=1\,,
&
\nonumber\\
&
Z_{V}=1+\epsilon\,a_{14}\,,\qquad
Z_{c}^{1/2}=1-\epsilon\,a_{16}\,,\qquad
Z_{\psi}^{1/2}=1+\epsilon\,a_{18}\,,
&
\nonumber\\
&\displaystyle
Z^{ab}_{\xi}=\delta^{ab}+\epsilon\,f_{1}^{ab}(\xi)\,,\qquad
Z^{ab}_{K}=\delta^{ab}-\epsilon\,\bigg(\frac{\partial f_{1}^{cb}(\xi)}{\partial \xi^{a}}\xi^{c}+f_{1}^{ab}(\xi)\bigg)\,,
&
\nonumber\\
&
Z_{J}=1+\epsilon\,a_{12}\,,\qquad
Z_{J,J_{\psi}}=\epsilon\,a_{13}\,,\qquad
Z_{J_{\psi}}=1+\epsilon\,a_{1}\,,
&
\nonumber\\
&
Z_{\zeta}=1+\epsilon\,(a_{7}+2a_{12})\,,\qquad
Z_{\zeta'}=1+\epsilon\,(a_{8}+2a_1+a_{12})\,,\qquad
Z_{\zeta''}=1+\epsilon(a_{9}+4a_{1})
&
\end{eqnarray}
and
\begin{eqnarray}
&
Z^{1/2}_{\tau}=Z_{\mathcal{J}}=Z^{-1}_{\Xi}=Z^{1/2}_{\bar{\eta}}\,,
&
\nonumber\\
&
Z_{\alpha}^{1/2}=Z_{b}^{-1/2}=Z^{-1/2}_{\bar{c}}=Z_{\Omega}^{-1}=Z^{1/2}_{A}\,,
&
\nonumber\\
&
Z^{1/2}_{\bar{\psi}}=Z^{-1}_{U}=Z^{-1}_{\bar{U}}=Z^{1/2}_{\psi}\,,
&
\nonumber\\
&
Z_{\bar{V}}=Z_{V}\,,\qquad
Z_{\zeta'\zeta}=Z_{\zeta''\zeta'}=Z_{J,J_{\psi}}\,.
&
\end{eqnarray}

\noindent One sees thus that the whole counterterm can be reabsorved into the starting action $\Sigma$. This concludes the proof of the renormalizability of the action $\Sigma$ to all orders of  perturbation theory. As usual, see also \cite{Fiorentini:2016rwx,Capri:2017bfd}, the auxiliary Stueckelberg field $\xi^a$, being dimensionless, renormalizes in a nonlinear way through the quantity $f_{1}^{ab}(\xi)$, which is a power series in $\xi^a$.  Taking finally the physical values of the sources $J$ and $J_{\psi}$, eq.\eqref{phys_values}, we obtain the renormalization of the corresponding mass parameters, namely 
\begin{eqnarray}
m^{2}_{0}&=&m^{2}+\epsilon\,(a_{12}\,m^{2}+a_{13}\,m^{2}_{\psi})\,,\nonumber\\
m_{\psi,0}&=&m_{\psi}+\epsilon\,a_{1}\,m_{\psi}\,.
\end{eqnarray}
As one can observe, there is a mixing between the mass parameters in the renormalization process.  This suggests that even if we do not include the gauge invariant operator  $A^{h}_\mu A^{h}_\mu $ from the beginning, {\it i.e.} $(m^{2}=0)$, it would show up through  quantum corrections if the theory contains a fermionic mass parameter like $m_{\psi}$.

\section{Conclusions}
\label{concl}

\noindent In the present  work, we have pursued the previous investigation started in  \cite{Fiorentini:2016rwx,Capri:2017bfd} by introducing, in addition of the gauge invariant composite fields $A^h_\mu$ and $A^h_\mu A^h_\mu$,  their spinor gauge invariant counterparts $(\psi^h, {\bar \psi}^h)$, eq.\eqref{local_psih}. \\

\noindent The main result obtained is that the starting action $\Sigma$ in presence of the gauge invariant composite operators $(\psi^h, {\bar \psi}^h)$ is renormalizable to all orders in perturbation theory. \\

\noindent As  underlined in the introduction, the gauge invariant composite operators $(\psi^h, {\bar \psi}^h)$ might give rise to several applications of interest. \\

\noindent The operators $(\psi^h, {\bar \psi}^h)$ could be employed in order to construct an effective kind of matter horizon function, in analogy with the so-called Gribov-Zwanziger horizon function \cite{Vandersickel:2012tz} enabling to restrict the functional integral to the Gribov region $\Omega$, to get rid of the Gribov copies. More precisely, the operators $(\psi^h, {\bar \psi}^h)$ allow us to write down the following non-local term 
\begin{eqnarray}
H_{\psi}(A^{h},\psi^{h})
  =  g^{2}\int d^{4}xd^{4}y\,(\bar{\psi}^{h})_{\alpha}^{i}(x)T^{a,ij}\left\{- [\partial_\mu D_\mu(A^{h})]^{-1}\right\}^{ab}\!(x,y)\,T^{b,jk}(\psi^{h})_{\alpha}^{k}(y)\,. 
  \label{horizon}
\end{eqnarray}
Expression \eqref{horizon} provides a manifest BRST invariant generalization of the term already introduced in \cite{Capri:2014bsa}. In particular, once including \eqref{horizon}
in the starting action and after a suitable localization procedure \cite{Capri:2014bsa}, it yields an analytic form for the quark propagator which is in qualitative agreement with the available lattice data in the Landau gauge \cite{Parappilly:2005ei}. Moreover, the manifest BRST invariance of expression \eqref{horizon} would enable us to study the so-called effective momentum dependent quark mass function ${\mathcal M}(p^2)$, see for example \cite{Capri:2014bsa,Capri:2017abz,Parappilly:2005ei}, in a BRST invariant way. \\

\noindent Another potential application of the operators $(\psi^h, {\bar \psi}^h)$ is that of allowing for a renormalizable non-Abelian Landau-Khalatnikov-Fradkin (LKF) transformations which include spinor fields, within the setup worked out recently in \cite{DeMeerleer:2018txc}. In fact, due to the BRST invariance of $(A^h_\mu,\psi^h, {\bar \psi}^h)$, it follows that the correlation functions $\langle A^h_{\mu_{1}}(x_1)...A^h_{\mu_{i}}(x_i) \psi^h(y_{1})....{\bar \psi}^h(y_{j})\rangle $ are independent from the gauge parameter $\alpha$, namely 
\begin{equation}
  \langle A^h_{\mu_{1}}(x_1)...A^h_{\mu_{i}}(x_i) \psi^h(y_{1})....{\bar \psi}^h(y_{j})\rangle_{\alpha \neq 0} =   \langle A^h_{\mu_{1}}(x_1)...A^h_{\mu_{i}}(x_i) \psi^h(y_{1})....{\bar \psi}^h(y_{j})\rangle_{\alpha = 0} \;. \label{lkfa}
\end{equation}
Once expanded in powers of the Stueckelberg field $\xi$, equation \eqref{lkfa} enables one to evaluate the Green function $\langle A_{\mu_{1}}(x_1)...A_{\mu_{i}}(x_i) \psi(y_{1})....{\bar \psi}(y_{j})\rangle_{\alpha \neq 0} $ in a given $\alpha$-gauge, with $\alpha \neq 0$, from the knowledge of the corresponding Green function evaluated in the Landau gauge, $\alpha=0$, yielding thus the LKF transformations within a renormalizable environment. \\\\ Finally, let us point out that equations \eqref{lkfa} might also be employed to sheed some light on nonperturbative aspects of gauge theories, like the infrared behaviour of the gauge and fermion propagators. As such, those equations could be exploited in order to show the gauge-independence of quantities like the chiral condensate, an issue of great relevance for the actual researches done in this field.

\section*{Acknowledgments}
\noindent This study was financed in part by the Coordena\c{c}\~{a}o de Aperfei\c{c}oamento de Pessoal de N\'{i}vel
Superior - Brasil (Capes) - Finance Code 001. The Conselho Nacional de Desenvolvimento Cient\'{\i}fico e Tecnol\'{o}gico (CNPq-Brazil), the Funda\c{c}\~{a}o de Amparo \`{a} Pesquisa do Estado do Rio de Janeiro (FAPERJ) and the SR2-UERJ are gratefully acknowledged for financial support. S. P. Sorella is a level PQ-1 researcher
under the program Produtividade em Pesquisa-CNPq, 300698/2009-7; M. A. L. Capri is a level PQ-2 researcher under
the program Produtividade em Pesquisa-CNPq, 302040/2017-0.


\appendix
\section{Discrete Symmetries of the Dirac Theory}

\label{CPT}

In addition to  the continuous Ward Identities described
in section \eqref{ward_ids}, there are three other symmetries which
helped us to constrain the final counterterm \eqref{parametric_form}.
The so-called $CPT$ symmetries: specified as parity, time-reversal
and charge conjugation. These symmetries enabled us to restrict a large set of
fermionic terms. In the next subsections, we will present in details, following \cite{Peskin:1995ev}, their action on 
the fermionic bilinears so as to render our renormalization procedure as much clear 
as possible \footnote{In order to be most detailed as possible, in this appendix we will
work with all the possible Dirac bilinear fields, such as $\bar{\psi}\psi$,
$\bar{\psi}\gamma^{\mu}\psi$, $i\bar{\psi}[\gamma^{\mu},\gamma^{\nu}]\psi$,
$\bar{\psi}\gamma^{\mu}\gamma^{5}\psi$, and $\bar{\psi}\gamma^{5}\psi$.}. 

\subsection{Parity}

\noindent Parity is a space-time symmetry, which assigns $(x_{4},x_{i})\rightarrow(x_{4},-x_{i})$,
where $i=1,2,3$. Therefore, to establish
the transformations laws under parity $(P)$ for each Dirac bilinear
fields, one has the transformations for $\psi$ and $\bar{\psi}$
characterized as 
\begin{eqnarray}
P\psi(x_{4},x_{i})P & = & -i(\gamma)_{4}\psi(x_{4},-x_{i})\,,\\
P\bar{\psi}(x_{4},x_{i})P & = & iP\psi^{\dagger}(x_{4},x_{i})P\gamma_{4}=i(P\psi(x_{4},x_{i})P)^{\dagger}\gamma_{4}=i\bar{\psi}(x_{4},-x_{i})\gamma_{4}\,.
\end{eqnarray}
 For the fermionic gauge invariant quantities $\psi^{h}$ and $\bar{\psi}^{h}$,
one has the same symmetries, namely 
\begin{eqnarray}
P\psi^{h}(x_{4},x_{i})P & = & -i\gamma_{4}\psi^{h}(x_{4},-x_{i})\,,\\
P\bar{\psi}^{h}(x_{4},x_{i})P & = & i\bar{\psi}^{h}(x_{4},-x_{i})\gamma_{4}\,.
\end{eqnarray}
Then the scalar bilinear transforms as 
\begin{eqnarray}
P\bar{\psi}\psi P & = & \bar{\psi}(x_{4},-x_{i})(i\gamma_{4})(-i\gamma_{4})\psi(x_{4},-x_{i})=\bar{\psi}\psi(x_{4},-x_{i}),
\end{eqnarray}
while for the vector one, we have  
\begin{eqnarray}
P\bar{\psi}\gamma_{\mu}\psi P & = & \bar{\psi}(i\gamma_{4})\gamma_{\mu}(-i\gamma_{4})\psi(x_{4},-x_{i})=\begin{cases}
\bar{\psi}\gamma_{\mu}\psi(x_{4},-x_{i}), & \mu=4;\\
-\bar{\psi}\gamma_{\mu}\psi(x_{4},-x_{i}), & \mu=1,2,3.
\end{cases}
\end{eqnarray}
Here, the vector displays the same minus sign on the spatial components
as does the space-time vector $x^{\mu}$. Analogously, the symmetries
of the pseudo-scalar and pseudo-vector \footnote{Both bilinears have an extra negative sign in the parity transformation.}
are given by 
\begin{eqnarray}
Pi\bar{\psi}\gamma_{5}\psi P & = & i\bar{\psi}(i\gamma_{4})\gamma_{5}(-i\gamma_{4})\psi(x_{4},-x_{i})=-i\bar{\psi}\gamma_{5}\psi(x_{4},-x_{i})\,,
\end{eqnarray}
\begin{eqnarray}
P\bar{\psi}\gamma_{\mu}\gamma_{5}\psi P & = & \bar{\psi}(i\gamma_{4})\gamma_{\mu}\gamma_{5}(-i\gamma_{4})\psi(x_{4},-x_{i})=\begin{cases}
-\bar{\psi}\gamma_{\mu}\gamma_{5}\psi, & \mu=4;\\
+\bar{\psi}\gamma_{\mu}\gamma_{5}\psi, & \mu=1,2,3.
\end{cases}
\end{eqnarray}
It is important to remember that the word ``pseudo'' ensures an
extra minus sign in the parity symmetry case. Therefore, to study
the equation \eqref{bilinear} we used the following bilinear quantity 
\begin{eqnarray}
\bar{\psi}\left[\gamma_{\mu},\gamma_{\nu}\right]\psi & = & 2\bar{\psi}\sigma_{\mu\nu}\psi \;, 
\end{eqnarray}
where under parity symmetry, we have 
\begin{eqnarray}
iP\bar{\psi}\left[\gamma_{\mu},\gamma_{\nu}\right]\psi P & = & i\bar{\psi}(x_{4},-x_{i})\gamma_{4}\left[\gamma_{\mu},\gamma_{\nu}\right]\gamma_{4}\psi(x_{4},-x_{i})\,.
\end{eqnarray}
Using the commutation relations for the gamma matrices: $\gamma_{4}\left[\gamma_{4},\gamma_{i}\right]\gamma_{4}=-\left[\gamma_{4},\gamma_{i}\right]$,
$\gamma_{4}\left[\gamma_{i},\gamma_{j}\right]\gamma_{4}=\left[\gamma_{i},\gamma_{j}\right]$and
$\sigma_{\mu\nu}=\frac{1}{2}\left[\gamma_{\mu},\gamma_{\nu}\right]$,
one gets 
\begin{eqnarray}
iP\bar{\psi}(x_{4},x_{i})\sigma_{\mu\nu}\psi(x_{4},x_{i})P & = & \begin{cases}
-\bar{\psi}(x_{4},-x_{i})\sigma_{4j}\psi(x_{4},-x_{i}), & \mu=i=4,\nu=j=1,2,3;\\
\bar{\psi}(x_{4},-x_{i})\sigma_{ij}\psi(x_{4},-x_{i}) & \mu=i,\nu=j=1,2,3.
\end{cases}
\end{eqnarray}

\subsection{Time-Reversal}

\noindent The time-reversal is a discrete symmetry which acts as $\psi(x_{4},x_{i})$ $\rightarrow $  $\psi(-x_{4},x_{i})$,
with $i=1,2,3$, where the time Euclidean time component has been identified with $x_4$. The time-reversal transformations for  the Dirac and the fermionic gauge invariant fields are: 
\begin{eqnarray}
T\psi(x_{4},x_{i})T & = & (-\gamma_{1}\gamma_{3})\psi(-x_{4},x_{i})\,,
\end{eqnarray}
\begin{eqnarray}
T\bar{\psi}T & = & (T\psi T)^{\dagger}(-i\gamma_{4})^{*}=\psi^{\dagger}(-x_{4},x_{i})(-\gamma_{1}\gamma_{3})^{\dagger}(i\gamma_{4})=\bar{\psi}(-x_{4},x_{i})(\gamma_{1}\gamma_{3}),
\end{eqnarray}
\begin{eqnarray}
T\psi^{h}(x_{4},x_{i})T & = & (-\gamma_{1}\gamma_{3})\psi^{h}(-x_{4},x_{i})\,,
\end{eqnarray}
\begin{eqnarray}
T\bar{\psi}^{h}T & = & (T\psi^{h}T)^{\dagger}(-i\gamma_{4})^{*}=\bar{\psi}^{h}(-x_{4},x_{i})(\gamma_{1}\gamma_{3}).
\end{eqnarray}
The transformation law for the scalar bilinear is described
by 
\begin{eqnarray}
T\bar{\psi}\psi(x_{4},x_{i})T & = & \bar{\psi}(\gamma_{1}\gamma_{3})(-\gamma_{1}\gamma_{3})\psi(-x_{4},x_{i})=\bar{\psi}\psi(-x_{4},x_{i})\;, 
\end{eqnarray}
while for the  pseudo-scalar one gets: 
\begin{eqnarray}
Ti\bar{\psi}\gamma_{5}\psi T & = & -i\bar{\psi}(\gamma_{1}\gamma_{3})\gamma_{5}(-\gamma_{1}\gamma_{3})\psi(-x_{4},x_{i}).
\end{eqnarray}
For the vector, one has 
\begin{eqnarray}
T\bar{\psi}\gamma_{\mu}\psi T & = & \bar{\psi}(\gamma_{1}\gamma_{3})(\gamma_{\mu})^{*}(-\gamma_{1}\gamma_{3})\psi=\begin{cases}
\bar{\psi}\gamma_{\mu}\psi(-x_{4},x_{i}), & \mu=4;\\
-\bar{\psi}\gamma_{\mu}\psi(-x_{4},x_{i}), & \mu=1,2,3.
\end{cases}
\end{eqnarray}
Under time-reversal the pseudo-vector has the same transformation
of the vector, given by 
\begin{eqnarray}
T\bar{\psi}\gamma_{5}\psi T & = & \bar{\psi}(\gamma_{1}\gamma_{3})(\gamma_{5})^{*}(-\gamma_{1}\gamma_{3})\psi=\begin{cases}
\bar{\psi}\gamma_{5}\psi(-x_{4},x_{i}), & \mu=4,\\
-\bar{\psi}\gamma_{5}\psi(-x_{4},x_{i}) & \mu=1,2,3.
\end{cases}
\end{eqnarray}
The case $\bar{\psi}\left[\gamma_{\mu},\gamma_{\nu}\right]\psi=2\bar{\psi}\sigma_{\mu\nu}\psi$
is characterized by
\begin{eqnarray}
T\bar{\psi}\sigma_{\mu\nu}\psi T & = & \frac{1}{2}\bar{\psi}(-x_{4},x_{i})(\gamma_{1}\gamma_{3})\left[\gamma_{\mu},\gamma_{\nu}\right]^{*}(-\gamma_{1}\gamma_{3})\psi(-x_{4},x_{i})\nonumber \\
 & = & \begin{cases}
\bar{\psi}(-x_{4},x_{i})\sigma_{0j}\psi(-x_{4},x_{i}), & \mu=0,\nu=j=1,2,3;\\
-\bar{\psi}(-x_{4},x_{i})\sigma_{ij}\psi(-x_{4},x_{i}) & \mu=i,\nu=j=1,2,3.
\end{cases}
\end{eqnarray}

\subsection{Charge Conjugation}
\noindent Finally, let us present the charge conjugation $C$, which  acts on the  the Dirac and fermionic gauge invariant fields in the following way
\begin{eqnarray}
C\psi(x)C & = & (-\bar{\psi}\gamma_{4}\gamma_{2})^{T}\,,
\end{eqnarray}
\begin{eqnarray}
C\bar{\psi}(x)C & = & C\psi^{\dagger}C(-i\gamma_{4})=(-i\gamma_{2}\psi)^{T}(-i\gamma_{4})\nonumber \\
&=&(-i(-i\gamma_{4})\gamma_{2}\psi)^{T}  = (-\gamma_{4}\gamma_{2}\psi)^{T}
\end{eqnarray}
\begin{eqnarray}
C\psi^{h}(x)C & = & (-\bar{\psi}^{h}\gamma_{4}\gamma_{2})^{T}\,,
\end{eqnarray}
\begin{eqnarray}
C\bar{\psi}^{h}(x)C & = & (-\gamma_{4}\gamma_{2}\psi^{h})^{T}\,.
\end{eqnarray}
Let us now consider the bilinears. For the scalar type, one has
\begin{eqnarray}
C\bar{\psi}\psi C & = & (-\gamma_{4}\gamma_{2}\psi)^{T}(-\bar{\psi}\gamma_{4}\gamma_{2})^{T}=(\gamma_{4})^{\alpha\beta}(\gamma_{2})_{\beta}^{\delta}\psi_{\delta}\bar{\psi}^{\vartheta}(\gamma_{4})_{\vartheta}^{\varsigma}(\gamma_{2})_{\varsigma\alpha}\nonumber \\
 & = & -\bar{\psi}^{\vartheta}(\gamma_{4})_{\vartheta}^{\varsigma}(\gamma_{2})_{\varsigma\alpha}(\gamma_{4})^{\alpha\beta}(\gamma_{2})_{\beta}^{\delta}\psi_{\delta}=\bar{\psi}\gamma_{2}\gamma_{4}\gamma_{4}\gamma_{2}\psi\nonumber \\
 & = & -\bar{\psi}\psi\,.
\end{eqnarray}
For the pseudo-scalar:  
\begin{eqnarray}
Ci\bar{\psi}\gamma^{5}\psi C & = & i(-i\gamma^{0}\gamma^{2}\psi)^{T}\gamma^{5}(-i\bar{\psi}\gamma^{0}\gamma^{2})^{T}=i\bar{\psi}\gamma^{5}\psi.
\end{eqnarray}
The $\gamma^{0}$and $\gamma^{2}$ are symmetric matrices  while
$\gamma^{1}$and $\gamma^{3}$ are antisymmetric. Thus for the vector and pseudo-vector one gets: 
\begin{eqnarray}
C\bar{\psi}\gamma^{\mu}\psi C & = & \bar{\psi}\gamma^{\mu}\psi;\\
C\bar{\psi}\gamma^{\mu}\gamma^{5}\psi C & = & -\bar{\psi}\gamma^{\mu}\gamma^{5}\psi.
\end{eqnarray}
For the bilinear $i\bar{\psi}\left[\gamma^{\mu},\gamma^{\nu}\right]\psi=2\bar{\psi}\sigma^{\mu\nu}\psi$, we get
\begin{eqnarray}
C\bar{\psi}\sigma_{\mu\nu}\psi C & = & \frac{1}{2}(-\gamma_{4}\gamma_{2}\psi)^{T} \sigma_{\mu\nu}(-\bar{\psi}\gamma_{4}\gamma_{2})^{T}=-\bar{\psi}\gamma_{4}\gamma_{2}
(\sigma_{\mu\nu})^{T}\gamma_{4}\gamma_{2}\psi,
\end{eqnarray}
using again the symmetry properties of the gamma matrices. Finally,
\begin{eqnarray}
C\bar{\psi}(x_{4},x_{i})\sigma_{\mu\nu}\psi(x_{4},x_{i})C & = & \bar{\psi}\sigma_{\mu\nu}\psi.
\end{eqnarray}

\end{document}